\newcommand{\be}{\begin{equation}}
\newcommand{\ee}{\end{equation}}
\newcommand{\bea}{\begin{eqnarray}}
\newcommand{\eea}{\end{eqnarray}}
\documentclass[a4paper,11pt,amsmath,amssymb,amsfonts,onecolumn,nofootinbib]{article}
\pdfoutput=1
\usepackage{jheppub}
\usepackage[T1]{fontenc}
\usepackage{graphicx}
\usepackage{dcolumn}
\usepackage{bm}
\usepackage{amsmath}
\usepackage{amssymb}
\usepackage{slashed}
\def\d{d\kern-.8 ex\vrule height 1.3 ex depth-1.24 ex width .7 ex \kern .15 ex}
\def\D{D\kern-1.7 ex\vrule height .87 ex depth-.8 ex width .7 ex \kern .95 ex}

\begin{document}
\title{Replicas, averaging and factorization in the IIB matrix model}

\author{Mihailo \v{C}ubrovi\'c}
\affiliation{Center for the Study of Complex Systems, Institute of Physics Belgrade, University of Belgrade, Serbia}
\emailAdd{cubrovic@ipb.ac.rs}

\date{\today}

\abstract{
We study the partition functions of multiple replicas (copies) of D-brane configurations in the type IIB (IKKT) matrix model. We consider the quenched regime, where small fluctuations of the matrices are superimposed onto the slow (quenched) dynamics of
the background, so the partition function is an ensemble average over the background. Interacting D-branes always factorize in a simple way. On the other hand, the non-interacting BPS configurations may or may not factorize depending on the number of
replicas, and their factorization mechanism is more involved as the corresponding saddle-point solutions (half-wormholes) break the replica symmetry. We argue that the simple factorization mechanism of interacting branes is actually more interesting as it
carries the specific signatures of quantum gravity, which are absent from disordered field theories like the SYK model.
}

\maketitle
\flushbottom

\section{Introduction}\label{sec1}

The factorization problem in gravity and holography has come into the spotlight with the discovery of replica wormholes, entanglement islands and the path they open toward a possible solution of the black hole information paradox
\cite{eastcoast,westcoast,review}. It also resonates with the general growth of knowledge in quantum information and its relation to black holes, scrambling \cite{scramble1,scramble2,scramble3} and quantum chaos \cite{thebook,randmat}. Stated simply, the 
puzzle lies in the fact that the spacetime (Euclidean)\footnote{Usually, the literature speaks of spacetime=Euclidean wormholes versus spatial=Lorentzian wormholes. But to be precise, Euclidean vs. Lorentzian signature is just a matter of choice, the true 
difference is between spatial wormholes which act as bridges of Kip Thorne style, and spacetime wormholes considered in this paper and in most works on factorization and averaging, which affect also the time direction. Therefore, even though the latter 
are usually considered in Euclidean time (also in this paper), one could also study them in Lorentzian signature. But one should still bear in mind that Euclidean/Lorentzian wormholes is the term often used in the literature, meaning really spacetime/
spatial wormholes.} wormholes constructed in \cite{eastcoast,westcoast}, which are crucial to save the Page curve and the unitarity of black hole evaporation, imply that the partition function of two copies (two replicas) of a gravitating system does not
equal the square of a single-replica partition function: $Z_\mathrm{1grav}^2\neq Z_\mathrm{2grav}$. But at least in asymptoticlly anti-de-Sitter (AdS) geometries this contradicts the field theory intuition that the partition function of two identical
decoupled systems should always factorize: $Z_\mathrm{1CFT}^2=Z_\mathrm{2CFT}$. The way out proposed in several works \cite{julian,belinsky,syknonaverage,wormnonaverage,wormquant} is that holography performs some kind of averaging or coarse-graning so that
the dual CFT partition function is really an expectation value over some distribution:
\be
\langle Z_\mathrm{1CFT}\rangle^2\neq\langle Z_\mathrm{2CFT}\rangle.
\ee
In this case, the nonfactorization ceases being a puzzle -- of course there is no reason that expectation values factorize. The averaging could in principle be carried over disorder ("explicit") or it could really be some kind of self-averaging in a
chaotic system, i.e. some kind of coarse-graning.

A prototypical framework for explicit averaging is the Sachdev-Ye-Kitaev (SYK) model, featuring a system of Majorana fermions with all-to-all coupling, the coupling strengths being a quenched random variable. In a certain 
regime, this system is dual to gravity in AdS${}_2$ \cite{sykads2}, making it perfect for studying the factorization problem. A simplified version -- SYK model in a single time point -- was analyzed in \cite{wormnonaverage,wormhalf} and the outcome is
very pleasing: although the wormhole configurations (which couple different replicas) are non-factorizing, there are additional solutions, dubbed half-wormholes, which restore the factorization; half-wormholes depend strongly on the choice of 
microscopic couplings. A similar picture was found to hold also in other systems like tensor models, random matrices and the two-site SYK model \cite{wormhalf2,wormblah,wormhalfgarcia,wormeuc}. In
\cite{wormprr,wormjap,wormams} the traversable spatial ("Lorentzian") wormholes have also been related to field theories averaged over the states or over the operators, starting, as could be expected, from thermofield-dynamics (TFD)-like states. Some
recent generalizations are found in \cite{wormpart,wormberry0,wormberry,wormberk} and in particular in \cite{gravfact}, where the authors find that nonlocally interacting bulk branes in 2D gravity provide a mechanism which restores factorization in absence
of any explicit disorder.

Our goal is to understand these workings of (half)wormholes in systems with an ensemble average but directly in quantum gravity, not in a field theory like SYK or a matrix model. The puzzle is now the following: is the averaging an operation which is
somehow "automatically" performed specifically by holography, or we can perform it in a gravitating system directly, without any reference to the dual field theory (or for gravitating systems which have no dual CFT at all)? We will take a quantum
gravity model (specifically the IIB string theory matrix model) and construct wormholes and half-wormholes by averaging over suitably chosen "quenched" degrees of freedom, thus repeating the logic of
\cite{julian,wormnonaverage,wormhalf,wormhalf2,wormblah,heckmann,tarekcoldhor} but directly in gravity. Therefore, we mainly aim to understand factorization as such, not specifically in the context of AdS/CFT. But we will also discuss how one can make
contact with AdS/CFT, by introducing a non-flat background metric; this is merely a first step and a full-fledged application of our findings to holography will be a subject of future work.

The arena for our work on wormholes and factorization is the model by Ishibashi, Kawai, Kitazawa and Tsuchiya (IKKT), proposed and developed in \cite{ikkt,ikkt2,ikktchep,ikktbig}. Like other matrix models of string and M theory (see \cite{uspekhi,ydri}
for a review), it is potentially capable of providing a nonperturbative description of string theory, including D-branes and other deep quantum effects, which are beyond the scope of old-style perturbative string theory. On the other hand, as a matrix
model, the IKKT system allows the explicit computation of observables and partition functions in a controlled way, and it has a lot in common with matrix models in field theory. In fact, in a certain limit the system we study can also be thought of as
ten-dimensional Yang-Mills field theory with a quenched background field configuration \cite{eguchikawai,ikktquench}. This makes our results relevant in principle also in the field theory context, and highlights that the formal workings of
(non)factorization of partition functions are in a sense quite technical and independent of many physical details of the system.

Another important point is that the IKKT model is well-defined both in Lorentzian and Euclidean signature. While the latter is more convenient when studying the landscape of saddle-point solutions, as we can define a partition function in terms of the
Euclidean action $S_E$ in the usual way as $Z=\int\exp(-S_E)$, the former is more frequently studied in the literature, as the amplitude $\mathcal{A}=\int\exp(-\imath S_L)$ of the Lorentzian action $S_L$ is always real; this is not the case for the
partition function when the fermionic excitations are turned on. For example, the Lorentzian dynamics was argued to explain the effective 3+1-dimensionality of spacetime and other cosmologically relevant issues in
\cite{ikkt4d,ikkt4dprl,ikktdyn,ikkt4dklink,ikktklink}. We have opted for the  Euclidean model so we can readily read off the free energy and determine which of the wormhole and half-wormhole solutions are thermodynamically prefered, but the reader should
bear in mind that this is just a matter of convenience (see also the first footnote).

With some hindsight, we can say that a picture rather similar to the wormhole/half-wormhole story in SYK and similar models will emerge here. We have not found a single case where the factorization of $\langle Z^n\rangle$ is not restored already at leading
order in perturbation theory for $n\geq 4$, although there are cases where the factorization is violated for $n=2$. In some cases the factorization is trivial (when the averaged value $\langle Z^n\rangle$ is at leading order just the product of $n$ copies
of the leading-order estimate for $\langle Z\rangle$), and sometimes it is nontrivial, in the sense that it cannot be written in terms of $\langle Z\rangle$ contributions only. This distinction is interesting and depends on the geometry of the D-brane
configuration. All of this is happening in (discretized) string theory, without any reference to holographic duality (although of course, one expects that the nonperturbative IKKT model implicitly knows about the duality).

As a final word of caution, one should bear in mind that more general wormhole configurations exist which are not necessarily all related to averaging, see, e.g. \cite{wormads,wormbraket,wormmaxfield,wormfree} and the recent insights of \cite{witten}.
There is actually a very general geometric perspective on wormhole solutions, which works also in quantum mechanics (not only field theory), resting on the symplectic structure of the Hamiltonian dynamics \cite{wormverlinde}; this likewise suggests
Euclidean wormholes to be more general than the averaging-induced configurations in this paper.

\subsection{The sharp question}

After all this talk, we are ready to formulate in a sharp way the main question of the paper -- if the fluctuations of the D-brane solutions to the IKKT model factorize when averaged over the background fields. Factorization means that $n$ replicas of the
system behave in the same way as $n$ independent copies. In precise language, this says that the averaged partition functions satisfy $\langle Z^n\rangle\sim\langle Z\rangle^n$. A related concept is self-averaging. Self-averaging means that the expectation
value of $\langle Z^n\rangle$ is "close" to a "typical" $Z^n$ value for some generic realization of the quenched variables (brane matrices). In precise language, denoting the quenched variables by $\lambda$, self-averaging is formulated as
\be
\langle Z^n(\lambda)\rangle\sim Z^n\left(\sqrt{\langle\lambda^2\rangle}\right).\label{selfaver}
\ee
The above means simply that the average over all $\lambda$ values produces the same result at leading order as the value for an average $\lambda$; the reason we take the square root of $\langle\lambda^2\rangle$ and not simply $\langle\lambda\rangle$ is
that the latter is often zero ($\lambda$ will often have a symmetric distribution).

The outline of the paper is the following. In Section \ref{sec2} we sum up the essential physics of D-branes in the type IIB matrix model and define the quenched approximation, setting the stage for the main work. Sections \ref{sec3} and \ref{sec4} contain
the core of the paper: we find saddle-point solutions of $Z^n$, $\langle Z^n\rangle$ and $\langle Z\rangle^n$ and discuss their factorization properties, first for a simple, single-D-string configuration and then for interacting D-strings. Section
\ref{sec5} sums up the conclusions. In Appendix \ref{secappb} we show that a technical simplification made in the paper (putting the background fermions to zero) does not lead to any qualitative changes in our results. In Appendix \ref{secappa} we
demonstrate another important technical point (that the conclusions do not depend qualitatively on whether the eigenvalue distribution is Gaussian or uniform). Finally, in Appendix \ref{secappc} we discuss how the reasoning of this paper could be applied
specifically to the factorization puzzle of holography, in asymptotically AdS backgrounds.

\section{Setup: D-brane configurations in the IKKT model}\label{sec2}

Let us start from the action of the type IIB matrix model, as found by Ishibashi, Kawai, Kitazawa and Tsuchiya \cite{ikkt,ikkt2} by discretizing the Schild action for IIB superstrings:
\be
\label{act}S=-\mathrm{Tr}\left(\frac{1}{4}\left[X_\mu,X_\nu\right]^2+\frac{1}{2}\bar{\Psi}_\alpha\Gamma^\mu\left[X_\mu,\Psi_\alpha\right]\right),
\ee
Here, $\mu=1,\ldots 10$ are the spacetime (target space) dimensions and $\alpha=1,\ldots 16$ counts the Majorana-Weyl fermions. The gamma matrices are then $16\times 16$ matrices. Both the scalars and the spinors are $N\times N$ Hermitian matrices. The
size $N$ is also dynamic, corresponding to the auxiliary field $g$ in the Schild action (see e.g. \cite{uspekhi}). However, we will always work with fixed and large $N$, assuming as usual that the partition function is strongly dominated by a single
saddle point at some $N$. For this reason there is no sum over $N$ in the partition functions throughout the paper.

The equations of motion follow from (\ref{act}):
\be
\label{eoms}\left[X^\mu,\left[X^\mu,X^\nu\right]\right]=0,~~\left[X^\mu,\left(\Gamma^\mu\Psi\right)_\alpha\right]=0.
\ee
We work in the Euclidean signature, hence the spacetime metric is $\eta_{\mu\nu}=\mathrm{diag}(1,\ldots 1)$ and we do not need to differentiate between up and down indices. The partition function is now given in the usual way 
\be
\label{partfun}Z=\sum_N\int D[X_\mu]\int D[\Psi_\alpha]\int D[\bar{\Psi}_\alpha]\exp\left(-S\right).
\ee
When only the bosonic degrees of freedom are excited, the above partition function is positive definite. If, however, the fermionic matrices $\Psi$ are also nonzero, then their contribution to the path integral (the Pfaffian) is complex in Euclidean
signature.\footnote{This is not a problem by itself. Indeed, there lies the mechanism of the $SO(10)$ symmetry breaking studied in \cite{ikktsymbreak1,ikktsymbreak2} in the context of the origin of the four-dimensionality of the spacetime.} Therefore, we
will need to be careful in interpreting the results when we turn on also the $\Psi$ fields.

Now let us remember how D${}_p$ branes show up in the matrix model. What follows is a resume of the crucial aspects of brane solutions from \cite{ikkt,uspekhi,ikktquench}. Type IIB string
theory admits D${}_p$ branes with $p$ odd, starting from $p=-1$, i.e. D-instantons. Taking D-instantons as elementary degrees of freedom, we can write any configuration of size $N$ as a superposition of $N$ D-instantons at coordinates $\lambda_j$,
$j=1,\ldots N$. For a general D${}_p$ brane, the BPS condition and equations of motion from the action (\ref{act}) lead to solutions $X_\mu=A_\mu$ for the bosonic part, with $A_\mu$ of the form:\footnote{From now on, $I$ always denotes the $N\times N$ unit
matrix. If we need the unit matrix of a different size $N'$, we will write it explicitly as $I_{N'\times N'}$.}
\be
\label{dpbrane}A_\mu=\left(\frac{L_1}{2\pi}q_1,\frac{L_2}{2\pi}k_1,\ldots\frac{L_{2i-1}}{2\pi}q_{(p+1)/2},\frac{L_{2i}}{2\pi}k_{(p+1)/2},0\ldots,0\right),~~[q_i,k_i]\equiv\omega_iI=\imath\frac{L_{2i-1}L_{2i}}{2\pi N^{2/(p+1)}}I.
\ee
Here $q_i$ and $k_i$ are random Hermitian matrices with the eigenvalues $\lambda^{(q_i)}_j$ and $\lambda^{(k_i)}_j$ (of course, $p+1\leq 10$), and $L_\mu$ ($\mu=1\ldots p+1$) are the compactification radii of the coordinates $X_\mu$. The commutators
$\omega_i$ have the meaning of $\hbar$, and $\omega_i\to 0$ corresponds to the classical limit --  in this case $q_i$ and $k_i$ are just any commuting Hermitian matrices, describing the moduli of the theory. Multi-brane configurations are described by
block-diagonal matrices $q_i,k_i$ with $M_p$ blocks of size $N\times N$ for $M_p$ branes D${}_p$. We will specialize to configurations of two D-strings as this suffices to discuss the factorization. The solution to the equations of motion (\ref{eoms})
which corresponds to two strings at distance $\ell$, with angle $2\theta$ between the strings, reads \cite{ikkt}:
\bea
&&A_0=\left(\begin{matrix}q & 0\\ 0 & q\end{matrix}\right),~~A_1=\left(\begin{matrix}p\cos\theta & 0\\ 0 & p\cos\theta\end{matrix}\right),
~~A_2=\left(\begin{matrix}\frac{\ell}{2} & 0\\ 0 & -\frac{\ell}{2}\end{matrix}\right),~~A_3=\left(\begin{matrix}-p\sin\theta & 0\\ 0 & p\sin\theta\end{matrix}\right)\nonumber\\
&&A_4=\ldots=A_{10}=0,~~[p,q]=\imath\frac{L^2}{2\pi N}I.\label{stringpair}
\eea
For $\theta=0$ ($\theta=\pi$) we get two parallel (antiparallel) strings, and for $\ell=0$ the strings cross. Parallel D-branes (including D-strings), with $\theta=0$, are a special case: they are BPS states with zero on-shell action. This will be
important as the properties and mechanisms of factorization differ between the BPS and the interacting case.

\subsection{The background-field regime and the quenched regime of the IKKT model}

In the semiclassical regime one can study the fluctuations $a_\mu$ around the classical geometry described by matrices $A_\mu$ from (\ref{dpbrane}) or (\ref{stringpair}):
\be
\label{quench}X_\mu=A_\mu+a_\mu,~~\Psi_\alpha=\psi_\alpha,
\ee
where we have taken into account that the background configuration of the fermions is $\Psi_\alpha=0$. This spoils the supersymmetry and non-renormalization results, but does not change the essence of our conclusions; indeed, in the large $N$ limit this is
physically justified as the fermions are large-$N$ suppressed anyway.\footnote{Another way to say this is that the fermionic background can be put to zero in the semiclassical limit. Our general formalism is not limited to the semiclassical regime, but 
the physical picture does become simpler in that case, as we can interpret the matrix entries as spacetime coordinates of the constituent D-instantons making the D-brane. Such approximation is also exploited for similar reasons in some of the original IKKT papers \cite{ikkt,ikktchep}.} In Appendix \ref{secappb} we redo the main calculation of the paper in the presence
of nonzero background $\Psi_\alpha$ and show that the outcome is essentially the same, though different in quantitative details of the solution. In the classical (commuting) regime, the matrices $A_\mu$ can be diagonalized simultaneously, their eigenvalues
having the role of locations of D-instantons. In applications to D-brane dynamics and other nonperturbative string theory phenomena, as well as gravity and the dynamics of spacetime dimensions, the natural viewpoint is the background field approach, i.e.
to integrate out the quantum fluctuations $a_\mu$, $\psi_\alpha$ and study the effective semiclassical action for the diagonal matrices $A_\mu$.

The opposite limit is the quenched IKKT model of \cite{ikktquench}, where the moduli $A_\mu$ are considered as fixed backgrounds and not integrated over. Now $a_\mu$ (and $\psi_\alpha$, if turned on) are "dynamical" variables, and $A_\mu$ are quenched, so
the partition function $Z$ is tied to a specific realization of $A_\mu$. This approach is natural if we want to study the quantum dynamics of perturbative string excitations in a given geometry. It can also be interpreted in the context of the Eguchi-Kawai
finding \cite{eguchikawai} that the Lagrangian of the Yang-Mills gauge theory in the large $N$ limit loses the derivative terms and reduces just to (\ref{act}). In that context the dynamics of $a_\mu$ and $\psi_\alpha$ is basically the dynamics of a
Yang-Mills system in the presence of quenched disorder.\footnote{The equivalence is not complete, as the quenched IKKT model contains zero modes, from the terms which couple $a^\mu$ and $\psi_\alpha$; such terms do not exist in the large-$N$ Yang-Mills
theory. However, at leading -- quadratic -- order such terms vanish also in the quenched IKKT model (see Eqs.~(\ref{act2}-\ref{act4})).} In other words, \emph{the fields $A_\mu$ play the same role as disorder in the SYK and similar models, describing on
one hand a prototypical strongly interacting large $N$ field theory and on the other hand string theory in deep nonperturbative regime}. This is our motivation to study the quenched IKKT model. Furthermore, AdS/CFT suggests that the background-field limit
and the quenched limit show different aspects of the same physics -- the Eguchi-Kawai system must have a dual description in terms of D-brane stacks. Therefore, the quenched IKKT model is a robust system of equations with several important
physical aspects and should provide us an excellent testing ground for the factorization puzzle. From now on, we will denote the non-averaged quantities by ordinary letters, e.g. $Z$ for the partition function, and the quantities averaged over the
realizations of the quenched variables $A_\mu$ will be denoted as expectation values, inside angular brackets, e.g. $\langle Z\rangle$.

Let us now write the IKKT action (\ref{act}) in terms of the background field $A_\mu$ and the fluctuations $a_\mu$, $\psi_\alpha$, as in (\ref{quench}). The quadratic ($S_2$) and quartic ($S_4$) contributions in $a_\mu$ and $\psi_\alpha$
read:
\bea
S&=&S_2+S_4\\
S_2&=&\mathrm{Tr}\left[-\frac{1}{4}a_\mu\left(P^2\delta_{\mu\nu}+2F_{\mu\nu}\right)a_\nu+\bar{\psi}_\alpha\slashed{P}\psi_\alpha-\bar{c}P^2c\right],\label{act2}\\
S_4&=&\mathrm{Tr}\left[\bar{\psi}_\alpha\slashed{a}\psi_\alpha-2\left(P_\mu a^\mu\right)\left(a^\nu a_\nu\right)-\frac{1}{2}\left[a^\mu,a^\nu\right]\left[a_\mu,a_\nu\right]\right],\label{act4}
\eea
where we have introduced the superoperators $P_\mu$ and $F_{\mu\nu}$, acting on matrices as
\be
P_\mu\equiv\left[A_\mu,\cdot\right],~~F_{\mu\nu}=\left[\left[A_\mu,A_\nu\right],\cdot\right],~~P^2=P_\mu P^\mu.\label{pfdefs}
\ee
We will also sometimes use the matrix $f_{\mu\nu}\equiv [A_\mu,A_\nu]$. Finally, $c$ in (\ref{act2}) is the ghost field arising from the Faddeev-Popov ghost action. BPS states have $f_{\mu\nu}=\mathrm{const.}\times I$ with $I$ being the unit matrix or,
equivalently, $F_{\mu\nu}=0$; in that case the on-shell action vanishes, one-loop quantum corrections are absent and the eigenvalues of $A_\mu$ remain uniformly distributed.

As we explained, for our purposes it is enough to focus on the bosonic sector and put the background fermionic matrices to zero: $\Psi_\alpha=0$. This is always a consistent (if not fully generic) solution of the equations of motion, and it simplifies the calculations
significantly. Bose-only backgrounds always give real action but they are obviously not protected by supersymmetry so even if $F_{\mu\nu}=0$ there will be a logarithmic attraction of the eigenvalues as there is no fermionic sector to cancel the determinant
from the bosonic path integral; however it remains (trivially) true that the solutions with $F_{\mu\nu}=0$ describe a non-interacting configuration (in the Eguchi-Kawai picture, the field strength is zero). In general, even in zero fermionic background
there are still nonzero fermionic fluctuations $\psi_\alpha$. As a special case, we can turn off these too: $\psi_\alpha=0$. In that case only the first term in (\ref{act2}) is nonzero, and in (\ref{act4}) the second and the third term remain nonzero. We
will consider this special case as a warmup but will always include also the fermionic fluctuations $\psi_\alpha$ in the end.

\section{Replicas and factorization for a single D-string}\label{sec3}

Now that we have set up the formal framework, we can explore the main question of the paper: how do the partition functions of D-branes in the quenched IKKT model factorize? From now on, we will specialize to D-strings as the whole story remains the same
for higher-dimensional branes. In this section we give a detailed calculation for a single bosonic D-string, so the reader can get acquainted with the basic algorithm on the simplest example. Afterwards it will be straightforward to redo the calculation
for the more interesting configurations of interacting strings.

\subsection{Single copy -- direct calculation}

As we know \cite{uspekhi}, the background fields $A_\mu$ representing a D-brane are described by random Hermitian matrices $p,q$ which, according to (\ref{dpbrane}), satisfy the commutation relation (putting $L_1=L_2\equiv L$):
\be
[p,q]=\omega=\frac{L^2}{2\pi N}I.
\ee
This is a BPS state (although we do not see the supersymmetry when we put $\Psi_\alpha=0$) with $f_{21}=-f_{12}=\omega$ and $F_{\mu\nu}=0$. The eigenvalues $\lambda_\mu^i$ ($\mu=1,2$, $i=1\ldots N$) are distributed in the interval $-L\leq\lambda_i\leq L$.
Unlike the typical choice in the IKKT model literature, where the compactification radii $L_\mu$ define the hard cutoffs of the distribution, we adopt -- solely for computational reasons -- a soft cutoff with the Gaussian distribution of eigenvalues
(Gaussian Unitary Ensemble) for $A_\mu$ so the width of the Gaussian equals $L_\mu$. Our choice is unusual but on one hand it is no less physical (indeed, a hard cutoff is more of an idealization than a continuous Gaussian tail) and on the other hand more
convenient for calculations. However, we emphasize that nothing changes qualitatively even when we return to hard cutoffs -- in Appendix \ref{secappa} we show this explicitly, redoing the calculations with hard cutoff regularization.

From (\ref{partfun}) we now have the following form for the partition function in the non-averaged and in the averaged form, respectively:
\bea
Z&=&\int D[a_\mu]e^{-S(a_\mu;A_\mu)}\\
\langle Z\rangle&=&\int D[A_\mu]\int D[a_\mu]e^{-S(a_\mu;A_\mu)}\mathcal{P}(A_\mu)=\int D[a_\mu]\int d^{2N}\lambda_\mu^ie^{-S\left(a_\mu;\lambda_\mu^i\right)-\frac{1}{2L^2}\lambda_\mu^i\lambda^{\mu i}}.~~~~~\label{zbasic}
\eea
In the second equality, we have expressed $A_\mu$ in terms of its eigenvalues $\lambda_\mu^i$ and likewise the Gaussian measure (regulator) $\mathcal{P}$ is written out explicitly in terms of the width of the distribution $L$, as
$-\lambda_\mu\lambda^\mu/2L^2$. Now we want to write out the action in (\ref{zbasic}) in terms of matrices $A_0=p$ and $A_1=q$. It is easiest to diagonalize the matrices and work in the eigenbases. Then we have
$A_\mu=\mathrm{diag}(\lambda^\mu_1,\ldots\lambda^\mu_N)$. The superoperators $P_\mu$ are now represented as\footnote{Superoperators act on $N\times N$ matrices, hence they are strictly speaking the Kronecker products of two $N\times N$ matrices. For
convenience, we write them simply as $N^2\times N^2$ matrices.}
\bea
\nonumber &&P_1=q\otimes I-I\otimes q,~~\left(P_1\right)^{ij}_{kl}=q_{ik}\delta_{jl}-q_{jl}\delta_{ik}\\
&&P_2=k\otimes I-I\otimes k,~~\left(P_2\right)^{ij}_{kl}=k_{ik}\delta_{jl}-k_{jl}\delta_{ik}~~~.\label{pmu}
\eea
However, since the canonical momentum and the conjugate coordinate do not commute, we cannot assume both matrices to be diagonal at the same time, i.e. in the same basis. The above representation (\ref{pmu}) takes each $P_\mu$ in its own eigenbasis. In
order to give the reader a clearer intuitive grasp, we can list a few supermatrix elements, say for $P_1$:
\bea
\nonumber\left(P_1\right)_{11}&=&
\left(\begin{matrix}0 & p_{12} & \ldots & p_{1N}\\ p_{21} & p_{22}-p_{11} & \ldots & p_{2N}\\ \ldots & \ldots & \ldots & \ldots \\ p_{N1} & p_{N2} & \ldots & p_{NN}-p_{11}\end{matrix}\right)\\
\nonumber\left(P_1\right)_{12}&=&\mathrm{diag}(-p_{12}, \ldots -p_{12}),~\left(P_1\right)_{1N}=\mathrm{diag}(-p_{1N}, \ldots -p_{1N})\\
\nonumber \ldots \\
\nonumber \left(P_1\right)_{21}&=&\mathrm{diag}(-p_{21}, \ldots -p_{21}),~
\left(P_1\right)_{22}=
\left(\begin{matrix}p_{11}-p_{22} & p_{12} & \ldots & p_{1N}\\ p_{21} & 0 & \ldots & p_{2N}\\ \ldots & \ldots & \ldots & \ldots \\ p_{N1} & p_{N2} & \ldots & p_{NN}-p_{22}\end{matrix}\right)\\
\nonumber\left(P_1\right)_{N,N-1}&=&\mathrm{diag}(-p_{N,N-1}, \ldots -p_{N,N-1})\\
\left(P_1\right)_{NN}&=&
\left(\begin{matrix}p_{11}-p_{NN} & p_{12} & \ldots & p_{1N}\\ p_{21} & p_{22}-p_{NN} & \ldots & p_{2N}\\ \ldots & \ldots & \ldots & \ldots \\ p_{N1} & \ldots & p_{N,N-1}-p_{NN} & 0\end{matrix}\right).~~~\label{pmularge}
\eea
Averaging over the quenched variables now requires either adopting a single basis and transforming all $P_\mu$ but one (say $P_1$) from the form (\ref{pmu}) to this fixed basis, or keeping each $P_\mu$ in its own eigenbasis -- but then we have to
divide the integral measure by the volume of the unitary matrices that transform each supermatrix to its eigenbasis. The latter is more convenient, and it gives rise to the eigenvalue attraction term. Plugging in the representation (\ref{pmu}) into $Z$ from
(\ref{zbasic}) and taking into account the basis change we get
\bea
\langle Z\rangle&=&\int D[a_\mu]\int d^{2N}\lambda_{\mu i}\Pi_{i<j}\left(\lambda_{\mu i}-\lambda_{\mu j}\right)^2
\nonumber\exp\left[-\frac{1}{4}a^\dagger_{\mu ij}\left(\lambda_{\mu i}^2+\lambda_{\mu j}^2\right)a_{\mu kl}\delta_{jk}\delta_{il}-\frac{1}{2L^2}\lambda_{\mu i}^2\right]=\\
\label{z1saddle}&=&\int D[a_\mu]e^{-W_1},~~W_1=\frac{1}{2}\sum_\mu\log\det\left(\frac{I}{L_\mu^2}+2a_\mu^\dagger a_\mu-2I\mathrm{Tr}a_\mu^\dagger a_\mu\right).
\eea
We have introduced the effective macroscopic action $W_1$ akin to the thermodynamic free energy. The first line in (\ref{z1saddle}) is obtained by inserting the matrix representations for $P$ into the general expression (\ref{zbasic}) taking into account
the basic change (see e.g. \cite{ikktchep}), and the final expression for $W_1$ in the second line comes from performing the Gaussian integral over $\lambda^\mu_i$ and the second-order expansion of the determinant in small $a^\dagger a$ and $1/L^{2N}$.
Since the partition function is not a power-law function of $a$ and $a^\dagger$, the series expansion has correlation functions $a^\dagger a\ldots a^\dagger a$ of arbitrarily high order. Their meaning is better grasped in the collective field formalism
that we introduce in the next subsection.

\subsection{Collective fields and replicas}

\subsubsection{Warmup: single partition function again}

The time is ripe to introduce the collective fields. We follow the formalism of \cite{wormnonaverage,wormhalf,wormblah} and largely adopt the notation of \cite{wormnonaverage}. The idea is the following: we define a bilinear operator $g$ as being equal to
the current $a^\dagger a$. We impose this equality as a constraint through the Dirac delta functional, and finally replace the $a^\dagger a$-dependent terms by the appropriate functionals of the bilinear:\footnote{For the sake of brevity we leave out the
spatial indices $\mu$, $\nu$ etc. from now on. We will only write them when leaving them out could lead to confusion.}
\bea
\nonumber&&\langle Z\rangle=\int D[a]\int D[g]\int d\lambda\exp\left[-\frac{1}{2}\mathrm{Tr}a^\dagger P^2a-\frac{2\left(\mathrm{Tr}g-\mathrm{Tr}a^\dagger a\right)}{L^{2N-2}}\right]
\delta\left(g-a^\dagger a\right)\mathcal{P}(\lambda)=\\
\nonumber &&=\int D[a]\int D[g]\int D[s]\int d\lambda
\exp\left[-\frac{1}{2}\mathrm{Tr}a^\dagger P^2a-\frac{2\left(\mathrm{Tr}g-\mathrm{Tr}a^\dagger a\right)}{L^{2N-2}}-\imath\mathrm{Tr}\left[s^\dagger\left(g-a^\dagger a\right)\right]\right]\mathcal{P}(\lambda)=\\
\nonumber &&=\int D[a]\int D[g]\int \frac{D[s]}{2^N\pi^{N^2}}
\exp\bigg[-\frac{1}{2}\log\det\bigg(\frac{I}{L^2}+2a^\dagger a-2I\mathrm{Tr}a^\dagger a\bigg)-\frac{2\big(\mathrm{Tr}g-\mathrm{Tr}a^\dagger a\big)}{L^{2N-2}}-\\
\label{z1coll} &&-\imath\mathrm{Tr}\left[s^\dagger\left(g-a^\dagger a\right)\right]\bigg]\sim
\int D[g]\int\frac{D[s]}{2^N\pi^{N^2}}\exp\left[-\frac{1}{2}\log\det s-\imath\mathrm{Tr}s^\dagger g-\frac{2}{L^{2N-2}}\mathrm{Tr}g\right].~~~
\eea
In the first line, we have inserted into the effective action the term proportional to $\mathrm{Tr}g-\mathrm{Tr}a^\dagger a$ which, at leading order in $a^\dagger a$, equals zero on-shell, i.e. when the constraint $g=a^\dagger a$ is
obeyed. In the second line, we have implemented the Dirac delta in the usual way, through the auxiliary field $s$, and in the third line we have averaged over the eigenvalues of the quenched coordinates $\lambda$. Notice how the
inserted "zero term" precisely cancels all $a^\dagger,a$ dependence after averaging, which was of course the reason to introduce it in the first place. Again, this only holds at leading order in the $a$-fields and $1/L^N$. This is
important -- all our calculations are perturbative (unlike the simpler SYK quantum mechanics where the collective fields can be introduced in an exact way); for example, the last line in (\ref{z1coll}) holds at the order
$O\left(\left(a^\dagger a\right)^4\right)+O(1/L^{4N})$. We will not write explicitly such higher-order remainders. Higher order terms could be taken into account perturbatively, by introducing additional collective fields and
eventually closing the series by expressing the $l+1$-st order terms in terms of the loop integrals over the $l$-th and lower order terms. We will do this explictily for the $\langle Z^2\rangle$ calculation, when it will be necessary
to get anything nontrivial. For now, let us stay at the lowest order.

Now we can look for the saddle-point solutions of the effective action $W_1$ defined by the last line in (\ref{z1coll}) in the same way as in (\ref{z1saddle}), through $W_1\equiv-\log\langle Z\rangle$:
\bea
\nonumber &&\frac{\partial W_1}{\partial g}=\frac{2}{L^{2N-2}}I+\imath s=0,~\frac{\partial W_1}{\partial s}=\imath g+\frac{1}{2}\left(s^{-1}\right)^T=0\\
\label{z1collsaddle}&&s=\frac{2\imath}{L^{2N-2}}I,~~g=\frac{L^{2N-2}}{4}I.
\eea
Here, $I$ is the $N\times N$ unit matrix as usual. The solution (\ref{z1collsaddle}) is unique.

The important conclusion is that the nonzero solution (\ref{z1collsaddle}) consists of scalar matrices which scale as $L^{2N-2}$, and inserting them into $W_1$ yields
\be
\label{w1onshell}W_1\vert_\mathrm{on-shell}\sim (N^2-N)\log L+\frac{N}{2}\log 2.
\ee
Here and in the future we routinely disregard the contributions which go to zero when $N\to\infty$ or $L\to\infty$; we will only write them in a few special occasions when we want to emphasize that some part of the partition function contributes only
negligible terms. One can check that the same value follows from the expression (\ref{z1saddle}) after we solve the saddle-point equation for $a_\mu$. We will compare this value with the on-shell value of the effective action for $\langle Z^2\rangle$ to
check for factorization. One final remark: there is always a sum over the spacetime coordinate $\mu$ (in general $\mu=1,\ldots 2p$, in our case $\mu=1,2$), hence $W_1$ from (\ref{w1onshell}) is really multipled by $2$ ($2p$ for a general D${}_p$ brane).
But we do not write this factor explicitly in order not to clutter the notation; our $W_1$ (and similar for the two- and four-replica actions $W_2$ and $W_4$ we are yet to compute) is really the effective action per spacetime dimension (and since we work
in Euclidean signature all dimensions are equivalent).

\subsubsection{Double and quadruple partition function}

Now that we have the averaged partition function $\langle Z\rangle$ for a D-brane, our question is: does the replicated partition function $\langle Z^2\rangle$ factorize? With two copies (replicas), the collective fields show their true meaning. The
replicated partition function is\footnote{In order to make the equations more compact we do not write out the replica indices in the measure of the integrals. Therefore, we write $\int D[g]$ for $\int D[g_{AA}]$, or $\int D[a]$ for $\int D[a_A]$ and
similar. We still write the replica indices in the integrands (unlike the spatial indices $\mu,\nu\ldots$ which we usually leave out.)}
\be
Z^2=\int D[a]\exp\left[-\mathrm{Tr}\left(2\lambda^i\left(a_A^\dagger a_A\right)\lambda^i+\frac{I}{2L^2}\lambda_i^2\right)\right],
\ee
where now we have the left and right replicas, denoted by the indices $A,B,\ldots$ taking values $L$ or $R$. The fun part is the averaged function $\langle Z^2\rangle$:
\bea
\nonumber\langle Z^2\rangle&=&\int D[a]\int D[g]\int\frac{D[s]}{2\pi}\int d\lambda e^{-\frac{1}{2}\mathrm{Tr}a_A^\dagger P^2a_A}
e^{-V\left(g_{AA}\right)+V\left(a_A^\dagger a_A\right)}e^{-\imath\mathrm{Tr}\left[s^\dagger_{AA}\left(g_{AA}-a_A^\dagger a_A\right)\right]}\mathcal{P}(\lambda)=\\
\label{z2coll}&=&\int D[g]\int\frac{D[s]}{2^N\pi^{N^2}}\exp\left[-\imath\mathrm{Tr}\left(s^\dagger_{AA}g_{AA}\right)-\frac{1}{2}\log\det s_{AA}-V\left(g_{AA}\right)\right]\equiv e^{-W_2}.
\eea
Analogously to what we did for $\langle Z\rangle$, the term with $-V\left(g_{AA}\right)+V\left(a_A^\dagger a_A\right)$ in the exponent is the multiplication by unity on-shell, and the $s_{AA}$ auxiliary fields implement the Dirac delta
functional. Up to fourth order, the interaction terms $V$ read (from performing the Gaussian integration over $\lambda_i$ and expanding the determinant):
\bea
\nonumber V&=&V_2+V_4\\
\nonumber V_2&=&\frac{2}{L^{2N-2}}\mathrm{Tr}(g_{LL}+g_{RR})\\
\label{w2coll}V_4&=&\frac{4}{L^{2N-4}}\mathrm{Tr}^2(g_{LL}+g_{RR})-\frac{4}{L^{2N-4}}\mathrm{Tr}\left(g_{LL}^2+g_{LL}g_{RR}+g_{RR}g_{LL}+g_{RR}^2\right).
\eea
Again, this is just the expansion of the averaged function $\langle Z^2\rangle$ to fourth order in $a_A$, i.e. to second order in $g_{AA}$; the full expansion of the $\log\det$ term is infinite. This is a
consequence of the initial action $S_\mathrm{eff}$ being quadratic in the fields, so the effective actions $W_n$ contain the $\log\det$ term. If we had a linear coupling to the sources of the form $Ja$, we would get the by now familiar 
wormholes with collective fields $g_{LR}$, as in the SYK model in \cite{syknonaverage,wormhalf,wormblah}. Now the saddle point equations from (\ref{w2coll}) are:
\bea
\label{z2collsaddle1}&&\frac{1}{2}\left(s_{AA}^{-1}\right)^T+\imath g_{AA}=0~\Rightarrow~s_{AA}=\frac{\imath}{2}\left(g_{AA}^T\right)^{-1}\\
\label{z2collsaddle2}&&\imath s_{AA}+\frac{2I}{L^{2N-2}}+\frac{8I}{L^{2N-4}}\mathrm{Tr}(g_{LL}+g_{RR})-\frac{8}{L^{2N-4}}(g_{LL}+g_{RR})^T=0.
\eea
While complicated at first glance, this system of equations has a high degree of symmetry. First, it is manifestly $L\leftrightarrow R$ invariant; second, it allows (though does not require) the maximally symmetric ansatz, where $s$ and $g$ are scalar matrices. 
Their matrix indices are $i,j\in\lbrace 1,\ldots N\rbrace$ (which we do not write explicitly) and $A,B\in\lbrace L,R\rbrace$. Therefore a scalar solution assumes (1) the $U(N)$ group of the matrix model is fully preserved (which is logical in the
large-$N$, random matrix regime that we study) and (2) full \emph{replica symmetry} is preserved. The latter can be broken, and later on we will discuss this possibility in detail; but for now let us look at the replica-symmetric solution. We first insert
$s$ from (\ref{z2collsaddle1}) into (\ref{z2collsaddle2}) and then take the trace of both sides. Denoting $\mathrm{Tr}g_{LL}=\mathrm{Tr}g_{RR}\equiv t$, we find
\be
\label{z2collsaddletr}t\sim -\frac{1}{16L^2}\pm\frac{L^{N-2}}{4\sqrt{2}\sqrt{N}}.
\ee
At leading order in large $N$ and $L$, the two traces are symmetric, behaving as $\mathrm{Tr}g_{LL}=\mathrm{Tr}g_{RR}\sim\pm L^{N-2}/\sqrt{N}$. Still assuming maximal symmetry, this gives the following solution:
\be
\label{z2collsaddle}s=\pm\frac{2\imath}{t}I\otimes E,~~g=\pm tI\otimes E.
\ee
Here, $E$ is the two-by-two unit matrix in the replica space; its indices are $A,B\in\lbrace L,R\rbrace$. From now on we use this Kronecker product notation for the collective fields with replica indices: a field $g_{AA}$ can be written in the form
$B\otimes C$, with $B$ being an $N\times N$ matrix and $C$ being an $n\times n$ matrix with $n$ the number of replicas.

The solutions scale as $L^{N-2}$ instead of $L^{2N-2}$ for the single replica solution (\ref{z1collsaddle}). This fact already suggests that the two-replica solution does not factorize. Inserting (\ref{z2collsaddle}) into
(\ref{w2coll}) we find (for either sign in (\ref{z2collsaddle})):
\be
\label{w2onshell}W_2\vert_\mathrm{on-shell}\sim 2N^2\log L+L^{2N}/2\pm (2N)^{3/2}L^{N+2}.
\ee
Comparing to (\ref{w1onshell}), we see that indeed \emph{the two-replica solution does not factorize:} $\langle Z^2\rangle\neq\langle Z\rangle^2$. The first term in (\ref{w2onshell}) precisely equals $2W_1$\footnote{Remember that all the solutions and
on-shell action values are calculated for large $N$ and $L$, meaning we disregard the terms which tend to zero as $N\to\infty$. Such higher-order corrections would likely spoil the exact ratio but this is expected.} but it is actually strongly subleading
compared to the other two terms, which behave entirely differently, as $L^N$ and $L^{2N}$. Therefore, we can spot a contribution equaling $2W_1$ but there is also a much larger contribution absent in $W_1$.

So the partition function $\langle Z^2\rangle$ is not factorizing. Is it self-averaging? To check this, we can again follow \cite{wormnonaverage} and compute the four-replica solution. Exploiting the definition of the non-averaged $Z^2$ from
(\ref{z2coll}) we write:
\bea
\nonumber Z^2&=&\int \frac{D[s]}{2^N\pi^{N^2}}\chi(s;g)\Phi(s)\\
\nonumber\chi(s;g)&=&\int D[g]\exp\left[-\imath\mathrm{Tr}\left(s^\dagger_{AA}g_{AA}\right)-V\left(g_{AA}\right)\right]\\
\label{z2colldef}\Phi(s)&=&\int D[a]\exp\left[\imath\mathrm{Tr}\left(s^\dagger_{AA}a^\dagger_Aa_A\right)-\frac{1}{2}\mathrm{Tr}a^\dagger_AP^2a_A+V\left(a^\dagger_Aa_A\right)\right].
\eea
Here we have separated the partition function into the non-averaging part $\chi(s;g)$ which just doubles when computing $\langle Z^4\rangle$ as it is independent of $a_A$, and the part $\Phi(s)$ with $a_A$-dependent integrand which gets nontrivial additional
correlations when $Z^4$ is averaged. Now the replica indices $A$, $B$ can take values $L$, $R$, and $A'$, $B'$ take values $L'$, $R'$. In addition to the two-replica fields, we now also have combinations of the form $LLR'R'$ and similar:\footnote{In
this equation, exceptionally, we write the indices in the integrand, e.g., $\int D[a_{A'}]$, in order to emphasize that we now have extra copies of the fields.}
\bea
\nonumber &&\langle\Phi^2(s)\rangle=\int D[a_A]\int D[a_{A'}]\int d^N\lambda\mathcal{P}(\lambda_i)\times\\
\nonumber &&\times\exp\left[\imath\mathrm{Tr}\left(s^\dagger_{AA}a^\dagger_Aa_A+s^\dagger_{A'A'}a^\dagger_{A'}a_{A'}-\frac{1}{2}a^\dagger_AP^2a_A-\frac{1}{2}a^\dagger_{A'}P^2a_{A'}\right)+
V\left(a^\dagger_Aa_A\right)+V\left(a^\dagger_{A'}a_{A'}\right)\right]=\\
\nonumber&&=\int D[a_A]\int D[a_{A'}]
\exp\left[\imath\mathrm{Tr}\left(s^\dagger_{AA}a^\dagger_Aa_A+s^\dagger_{A'A'}a^\dagger_{A'}a_{A'}\right)+V\left(a^\dagger_Aa_A\right)+V\left(a^\dagger_{A'}a_{A'}\right)\right]\times\\
\label{z4coll0}&&\times\exp\left[-\frac{1}{2}\log\det\left(\frac{I}{L^2}+2a^\dagger_Aa_A+2a^\dagger_{A'}a_{A'}-2I\mathrm{Tr}a^\dagger_Aa_A-2I\mathrm{Tr}a^\dagger_{A'}a_{A'}\right)\right].
\eea
The next step is to expand the argument of the logarithm, which results in the mixing of $L$, $R$ and $L'$, $R'$. Expanding to order four, we generate a new interaction potential, mixing all four replicas, and again resort to the 
trick of multiplying by unity and introducing the four-replica collective field $G_{AAB'B'}$, together with the Dirac delta constraint implemented by $S_{AA'BB'}$:
\bea
\nonumber&&\langle\Phi^2(s)\rangle=\int D[a_A]\int D[a_{A'}]\int D[S]\int D[G]\times\\
\nonumber&&\exp\mathrm{Tr}\left[\imath s^\dagger_{AA}a^\dagger_Aa_A+\imath s^\dagger_{A'A'}a^\dagger_{A'}a_{A'}+\imath S^\dagger_{AAB'B'}\left(a^\dagger_Aa_Aa^\dagger_{B'}a_{B'}-G_{AAB'B'}\right)\right]\times\\
\nonumber&&\times\exp
\left[V\left(a^\dagger_Aa_A\right)+V\left(a^\dagger_{A'}a_{A'}\right)-\mathcal{V}\left(g_{AA},g_{B'B'};G_{AAB'B'}\right)+\mathcal{V}\left(a^\dagger_Aa_A,a^\dagger_{B'}a_{B'}\right)\right]=\\
\label{z4coll}&&=\int D[S]\int D[G]e^{-\tilde{W}_4},
\eea
with the effective action that now depends also on the new collective fields $G_{AAB'B'}\equiv a_A^\dagger a_Aa_{B'}^\dagger a_{B'}$. From (\ref{z2colldef}), the total effective action is the sum of $\tilde{W}_4$ from the last
equation and the non-averaging contribution from $\chi(s;g)$:
\bea
\nonumber &&W_4=\imath\mathrm{Tr}\left(s^\dagger_{AA}g_{AA}\right)+V\left(g_{AA}\right)+\tilde{W}_4\\
\nonumber &&\tilde{W}_4=\frac{1}{2}\left(\log\det s_{AB}+\log\det s_{A'B'}+\log\det S_{AAB'B'}\right)-\imath\mathrm{Tr}\left(S^\dagger_{AAB'B'}G_{AAB'B'}\right)+\mathcal{V}\\
\label{w4coll}&&\mathcal{V}\left(g_{AA},g_{B'B'};G_{AAB'B'}\right)=\frac{8}{L^{2N-4}}\mathrm{Tr}g_{AA}\mathrm{Tr}g_{B'B'}-\frac{4}{L^{2N-4}}\mathrm{Tr}G_{AAB'B'}.
\eea
This is the main formal result of this section. We will now consider various saddle-point solutions of the effective action (\ref{w4coll}), and consider which of these may restore the factorization and which are self-averaging.

\subsection{Half-wormholes}

\begin{figure}[h]
(A)\includegraphics[width=0.4\textwidth]{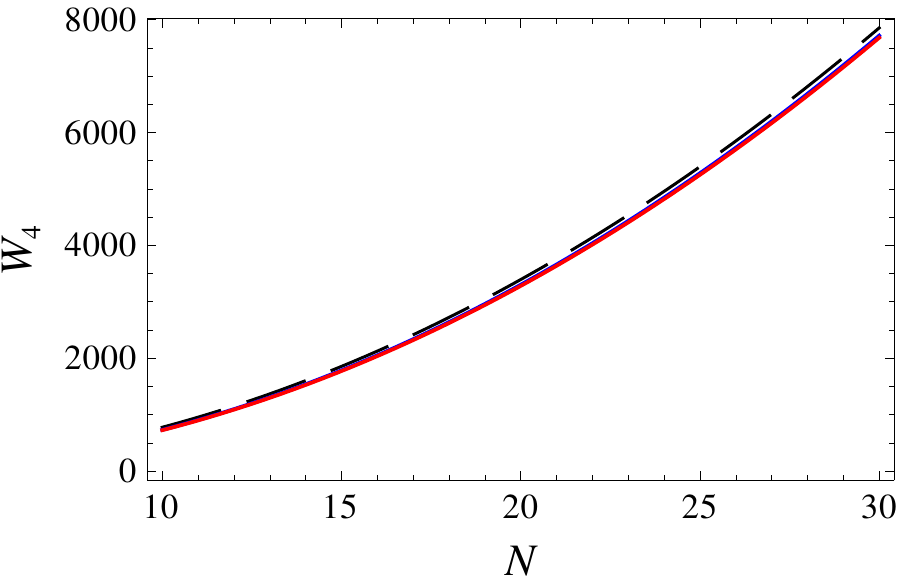}
(B)\includegraphics[width=0.4\textwidth]{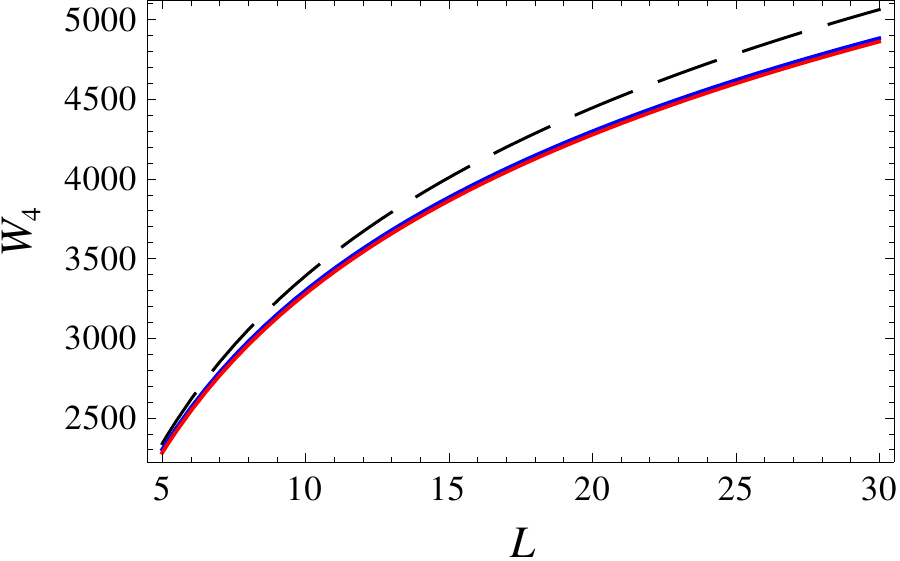}
\caption{\label{figfree} Effective action $W_4$ as a function of the the matrix size $N$ (A) and compactification radius $L$ (B) from analytical (red) and numerical (blue) calculations; the curves almost coincide. For 
reference we also show the single-replica solution $4W_1$ (black dashed) which is nearly equal to $W_4$, confirming the factorizing property.}
\end{figure}

The effective action (\ref{w4coll}) is slightly more involved but the highest symetry solutions are still easy to find. The saddle-point equations read:
\bea
\nonumber&&-\frac{1}{2}\left(g_{AA}^{-1}\right)^T+\frac{2I}{L^{2N-2}}+\frac{8I}{L^{2N-4}}\mathrm{Tr}(g_{LL}+g_{RR})-\frac{8}{L^{2N-4}}(g_{LL}+g_{RR})^T-\frac{8I}{L^{2N-4}}\mathrm{Tr}g_{B'B'}=0\\
\label{z4collsaddle}&&\frac{1}{2}\left(S_{AAB'B'}^{-1}\right)^T+\imath G_{AAB'B'}=0,~~-\imath S_{AAB'B'}-\frac{4}{L^{2N-4}}I=0.
\eea
Notice that the first and the second line are decoupled. The second line (the equations for $S$ and $G$) is linear and the solution is unique and left-right symmetric in the replica space (invariant to $L\leftrightarrow R$ and $L'\leftrightarrow R'$). In
the first line, we have already implemented the relation $\left(s_{AA}^{-1}\right)^T+2\imath g_{AA}=0$ which remains the same as in (\ref{z2collsaddle1}); that is why $s_{AA}$ is already elliminated from (\ref{z4collsaddle}). This equation is nonlinear and
potentially has many solutions. If we again take the fully replica-symmetric ansatz, we can easily solve for the trace as in (\ref{z2collsaddle1}-\ref{z2collsaddle2}). When everything is said and done the outcome is
\bea
&&s=\pm\frac{2\imath}{t}I\otimes E,~g=\pm tI\otimes E\\
&&S_{LLL'L'}=S_{LLR'R'}=S_{RRL'L'}=S_{RRR'R'}=\frac{4\imath}{L^{2N-4}}I\\
\label{solz4}&&G_{LLL'L'}=G_{LLR'R'}=G_{RRL'L'}=G_{RRR'R'}=\frac{L^{2N-4}}{8}I.\\
\label{w4onshell}&&W_4\vert_\mathrm{on-shell}\sim 4(N^2-N)\log L+2N\log 2-\sqrt{2N}/L^N.
\eea
The remaining components of $S$ and $G$ (those not listed in (\ref{solz4})) are zero. The first and second term in $W_4\vert_\mathrm{on-shell}$ come from $S$ and $G$ fields; the third term comes from $s$ and $g$. Although the first term clearly dominates
over the others, and the last term is negligible, we have deliberately written all three to emphasize the different contributions. 
We notice the following:
\begin{enumerate}
 \item The solution for $W_4$ factorizes at leading order, as we have $W_4\sim 4W_1$ on-shell (compare to (\ref{w1onshell})). What is more, it factorizes nontrivially, i.e. the expression for $W_4$ in (\ref{w4coll}) does not consist of four
 copies of $W_1$ (\ref{z1coll}). The leading contribution to (\ref{w4onshell}) comes roughly from the $\log\det S$ term in (\ref{w4coll}) and involves fields of the form $a^\dagger_La_La^\dagger_{L'}a_{L'}$ and similar, whereas the dominant contribution to
 $W_1$ comes from the term $\log\det s$ with fields of the form $a^\dagger_La_L$.
 \item There is no simple "wormhole" contribution, mixing the left and right copy, i.e. no fields $g_{LR}$, only $g_{LL}$ and $g_{RR}$. As we already said, this is so because there is no linear source coupling to $a$. However, since the fields $G_{AAB'B'}$
 (or explicitly $G_{LLR'R'}$ and $G_{RRL'L'}$) are nonzero, this solution can be called half-wormhole (HWH) in analogy with the terminology of \cite{syknonaverage}, already accepted in \cite{wormhalf,wormblah,wormhalfgarcia,wormjap}.\footnote{In fact, the
 name is not a very fortunate one, as there is nothing "halved" here; it a product of replicas just like a wormhole, only of different replicas. A more descriptive term would be "higher-order wormholes" but we will not attempt to change an already
 established name.} The half-wormhole thus dominates the action.
 \item The local contribution in replica space, the one coming from $g_{LL}$ and $g_{RR}$ fields, is negligible compared to the HWH part. It is also non-factorizing; but as in other cases in the literature, the HWH restores factorization.
 \item The solution is not self-averaging. To see this, we follow the criterion (\ref{selfaver}) and insert $\sqrt{\langle\lambda_\mu^2\rangle}=L_\mu$ into $Z$ as given in (\ref{zbasic}). Since now we deal with a free quadratic action (having fixed 
 $\lambda_\mu\mapsto L_\mu$) the $n$-replica partition function is easily found for any $n$: $W_n=nN\log L$; specifically, for $n=4$, we have $W_4\sim 4N\log L$. This does not capture the leading term in (\ref{w4onshell}).
\end{enumerate}
We can confirm our findings numerically by computing the full action for a numerically generated ensemble of quenched matrices $A_\mu$. In Fig.~\ref{figfree} we plot the numerical realization and the estimate (\ref{w4onshell}) in blue and red respectively.
We also plot the quadraple value of the single-replica action $W_1$ for comparison (black dashed). The analytical estimate is very good and the factorization near-perfect, as the black dashed curve almost falls on top of the blue and red one.

So while the action of the two-replica system $W_2$ does not factorize, the four-replica action $W_4$ has a HWH saddle point which restores factorization. In our view, the most interesting aspect of this is that the four-replica system \emph{factorizes
nontrivially}: the effective action (\ref{w4onshell}) is dominated by nonlocal terms, half-wormholes; it does not merely consist of four copies of $W_1$. Since our system contains explicit quenched disorder, it is no big surprise that in general 
$\langle Z^2\rangle\neq\langle Z\rangle^2$ (this is the same situation as for the SYK model). But the restoration of the factorization in the sense that $\langle Z^4\rangle=\langle Z\rangle^4$ was not \emph{a priori} expected, nor the fact that this
happens already in the collective fields formalism at leading order (disregarding an infinite series of higher-order terms in $\langle Z^4\rangle$). Even though we look directly at the (discretized) string theory action, which includes gravity, there is no
obvious geometric interpretation of the half-wormhole. This is not much of a surprise as the IKKT model should capture the nonperturbative stringy effects, far beyond general relativity and geometry.

\subsection{Fermionic contribution to D-string partition functions}

Now we turn on also the fermionic fluctuations $\psi_\alpha$ (while still putting the background fermions $\Psi_\alpha$ to zero). In addition to the collective fields introduced earlier, we now need also the fermion bilinear
$\gamma_{AB}=\bar{\psi}_A\psi_B$, with $A,B\in\lbrace L,R\rbrace$; in order to implement the constraint that defines $\gamma_{AB}$ we also have to introduce another auxiliary field called $\sigma_{AB}$ (analogous to $s_{AA}$). Actually, the fermionic part
may be more familiar to the reader as fermionic collective fields were investigated in the literature in much detail, mainly in the context of SYK model \cite{wormnonaverage,wormblah,wormprr}. We can first write $\langle Z\rangle$:
\bea
&&\langle Z\rangle=\int D[\psi]\int D[\bar{\psi}]\int D[a]\int D[g]\int D[\gamma]\int\frac{D[s]}{2^N\pi^{N^2}}\int\frac{D[\sigma]}{2^N\pi^{N^2}}\int d\lambda\exp\mathrm{Tr}\left(-\frac{1}{2}aP^2a-\bar{\psi}_\alpha\slashed{P}\psi_\alpha\right)
\times\nonumber\\
&&\times\exp
\left[-\imath\mathrm{Tr}\left(s^\dagger\left(g-a^\dagger a\right)+\sigma^\dagger\left(\gamma-\frac{1}{N}\bar{\psi}_\alpha\psi_\alpha\right)\right)-V(g)+V(a^\dagger a)-\Gamma(\gamma)+\Gamma\left(\frac{1}{N}\bar{\psi}_\alpha\psi_\alpha\right)\right]=
\nonumber\\
&&=\int D[g]\int D[\gamma]\int\frac{D[s]}{2^N\pi^{N^2}}\int\frac{D[\sigma]}{2^N\pi^{N^2}}e^{-\imath\mathrm{Tr}\left(s^\dagger g+\sigma^\dagger\gamma\right)-V(g)-\Gamma(\gamma)+\mathrm{Tr}\left(\gamma s^{-1}\gamma\right)+\log\det\sigma},\label{z1fermi}
\eea
where we have introduced the self-interaction potential $\Gamma(\gamma)=(L^2/4)\gamma^2$ for the collective fermionic fields analogously to $V(g)$ for the boson. At quadratic order $O(g^2+\gamma^2)$ there is no interaction between the Bose and Fermi
sectors. At the quartic level they couple, through the term $\gamma s^{-1}\gamma$ which comes from the term $\bar{\psi}\slashed{a}\psi$ in the original action. Defining the negative exponent in (\ref{z1fermi}) as $W_1$, we will now compare $W_1$ to the
four-replica effective action $W_4$. Let us first calculate $W_1$. The equations of motion read
\bea
&&\frac{2}{L^{2N-2}}I+\imath s=0,~\imath g+\frac{1}{2}\left(s^{-1}\right)^T+\gamma s^{-2}\gamma=0~\Rightarrow ~s=\frac{2\imath}{L^{2N-2}}I\otimes E,~g=\frac{L^{2N-2}}{4}I\otimes E\nonumber\\
&&\imath\sigma+\frac{L^2}{2}\gamma-\frac{1}{4}\left(\gamma s^{-1}+s^{-1}\gamma\right)=0,~\imath\gamma+\left(\sigma^{-1}\right)^T=0~\Rightarrow~\sigma=\imath\frac{L^{N-1}}{\sqrt{2}}I,~\gamma=\frac{\sqrt{2}}{L^{N-1}}I.~~~~~~~~~~~~\label{z1fermicoll}
\eea
This yields the same solution for the bosonic fields $g$ and $s$ as before, since their coupling to the fermions (the last term in the first line of (\ref{z1fermicoll})) only contributes a correction which goes to zero for $L,N\to\infty$; in fact, the 
$\gamma$-$s$ coupling does not influence the fermionic solution either at large $L,N$. The on-shell action is just the sum of the bosonic term (\ref{w1onshell}) and the fermionic term from (\ref{z1fermicoll}):
\be
W_1\vert_\mathrm{on-shell}=\left[(N^2-N)\log L+\frac{N}{2}\log 2\right]+\left[N\log L-\frac{N}{2}\log 2\right]=N^2\log L.\label{z1fermisol}
\ee
Now we derive the four-replica action $W_4$. In order to do this, we need to write first the two-replica function. Just like for bosons, it separates into the averaging and non-averaging part:
\bea
\nonumber Z^2&=&\int\frac{D[s]}{2^N\pi^{N^2}}\int\frac{D[\sigma]}{2^N\pi^{N^2}}\chi(s,\sigma;g,\gamma)\Phi(s,\sigma)\\
\nonumber\chi(s,\sigma;g,\gamma)&=&\int D[g]\int D[\gamma]\exp\left[-\imath\mathrm{Tr}\left(s^\dagger_{AA}g_{AA}+\sigma^\dagger_{AB}\gamma_{AB}\right)-V\left(g_{AA}\right)-\Gamma(\gamma_{AB})\right]\\
\Phi(s,\sigma)&=&\int D[a]\int D[\psi_A]\int D[\bar{\psi}_A]\exp\left[\imath\mathrm{Tr}\left(s^\dagger_{AA}a^\dagger_Aa_A\right)-\frac{1}{2}\mathrm{Tr}a^\dagger_AP^2a_A+V\left(a^\dagger_Aa_A\right)\right]\times\nonumber\\
&\times&\exp\left[\imath\mathrm{Tr}\left(\sigma^\dagger_{AA}\frac{\bar{\psi}_A\psi_A}{N}\right)-\mathrm{Tr}\bar{\psi}_A\slashed{P}\psi_A+\Gamma\left(\frac{\bar{\psi}_A\psi_A}{N}\right)-\mathrm{Tr}\bar{\psi}_A\slashed{a}_A\psi_A\right].
\eea
In order to test the factorization property we find $\langle\Phi^2\rangle$ through the steps similar to those in Eq.~(\ref{z4coll0}).
\bea
\nonumber&&\langle\Phi^2(s)\rangle=\int D[a_A]\int D[a_{A'}]\int d\bar{\psi}_A\int d\psi_A\int d\bar{\psi}_{B'}\int d\psi_{B'}\int\frac{dS}{2^N\pi^{N^2}}\int dG\times\\
\nonumber&&\times\exp
\left[\imath\mathrm{Tr}\left(s^\dagger_{AA}a^\dagger_Aa_A+s^\dagger_{A'A'}a^\dagger_{A'}a_{A'}+S^\dagger_{AAB'B'}a^\dagger_Aa_Aa^\dagger_{B'}a_{B'}+\sigma^\dagger_{AA}\frac{\bar{\psi}_A\psi_A}{N}\right)\right]\times\\
\nonumber&&\times\exp
\left[V\left(a^\dagger_Aa_A\right)+V\left(a^\dagger_{A'}a_{A'}\right)+\mathcal{V}\left(a^\dagger_Aa_A,a^\dagger_{B'}a_{B'}\right)+\Gamma\left(\frac{\bar{\psi}_A\psi_A}{N}\right)-\mathrm{Tr}\bar{\psi}_A\slashed{a}_A\psi_A\right]\times\\
\label{zfermicoll}&&\times\exp\left[-\imath\mathrm{Tr}\left(S^\dagger_{AAB'B'}G_{AAB'B'}\right)-\mathcal{V}\left(g_{AA},g_{B'B'};G_{AAB'B'}\right)\right]=\int\frac{dS}{2^N\pi^{N^2}}\int D[G]e^{-\tilde{W}_f}.~~~~~~~~~~
\eea
The final step is to integrate out the original fields. The bosonic integrals result in determinants and traces, as we have already seen, and the fermionic integrals are expressed in terms of Pfaffians, i.e. polynomials. Thanks to this the fermionic
bilinear can contain any combination of $L$, $L'$, $R$ and $R'$, giving rise to wormholes as we shall see. The effective action is finally:
\bea
\nonumber W_f&=&\imath\mathrm{Tr}\left(s^\dagger_{AA}g_{AA}+\sigma^\dagger_{AB}\gamma_{AB}\right)+V+\tilde{W}_f\\
\nonumber\tilde{W}_f&=&\frac{1}{2}\log\det\left(s_{AB}s_{A'B'}S_{AAB'B'}\right)-\imath\mathrm{Tr}\left(S^\dagger_{AAB'B'}G_{AAB'B'}\right)+\mathcal{V}+V_f\\
\label{wfermicoll}\mathcal{V}_f&=&
\frac{L^2}{4}\mathrm{Tr}\left(\gamma_{AB'}^2\right)-\mathrm{Tr}\left(\gamma_{AA}s_{AA}^{-1}\gamma_{AA}\right)-\log\mathrm{Tr}\left(\sigma_{LR}\sigma_{L'R'}-\sigma_{LL'}\sigma_{RR'}+\sigma_{LR'}\sigma_{RL'}\right),~~~~~~~~~
\eea
and the bosonic potentials $V$ and $\mathcal{V}$ are given in (\ref{w2coll}) and (\ref{w4coll}) respectively. The saddle-point equations (\ref{z4collsaddle}) now get additional terms and two additional equations from fermionic
contributions:
\bea
\nonumber&&\frac{1}{2}\left(s_{AA}^{-1}\right)^T+\imath g_{AA}+\gamma_{AA}s_{AA}^{-2}\gamma_{AA}=0\\
\nonumber&&\imath s_{AA}+\frac{2I}{L^{2N-2}}+\frac{8I}{L^{2N-4}}\mathrm{Tr}(g_{LL}+g_{RR})-\frac{8}{L^{2N-4}}(g_{LL}+g_{RR})^T-\frac{8I}{L^{2N-4}}\mathrm{Tr}g_{B'B'}=0\\
\nonumber&&\frac{1}{2}\left(S_{AAB'B'}^{-1}\right)^T+\imath G_{AAB'B'}=0,~~-\imath S_{AAB'B'}-\frac{4}{L^{2N-4}}I=0\\
\nonumber&&\imath\sigma_{AB}+\frac{L^2}{2}\gamma_{AB}-\frac{1}{4}\lbrace\gamma_{AB},s_{AB}^{-1}\rbrace\delta_{AB}=0\\
\label{z4fermisaddle}&&\imath\gamma_{AB}-\sigma_{A'B'}\left(\sigma_{CD}\sigma_{C'D'}-\sigma_{CC'}\sigma_{DD'}+\sigma_{CD'}\sigma_{DC'}\right)^{-1}=0.
\eea
This Bose-Fermi saddle-point system has a rich structure. We notice the following properties:
\begin{enumerate}
 \item The collective field $\gamma_{AB}$ can now couple different replicas already at the quadratic level. This is simply a consequence of the fact that fermions couple linearly to the quenched bosonic degrees of freedom $A_\mu$. We would have obtained 
 the same situation for the bosonic fields if we coupled them linearly to a source. The Pfaffian obtained upon integrating out $\psi_\alpha$ fields has a combinatorial structure which allows the breaking of replica symmetry, i.e. some $\sigma_{AB}$ may
 be zero and some nonzero.
 \item One solution is obtained by setting $\sigma_{LR}=\sigma_{L'R'}\equiv\sigma\neq 0$ while all other components are zero. This yields\footnote{In all examples we consider, the solutions for the bosonic fields $g$, $s$, $G$, $S$ do not change at leading
 order in the presence of fermions, hence we do not write the expressions for them again.}
 \bea
 &&\sigma_{LR}=\sigma_{L'R'}=\frac{L^{N-1}}{\sqrt{2}},~\gamma_{LR}=\gamma_{L'R'}=-\imath\frac{\sqrt{2}}{L^{N-1}}\label{solz4fermi1}\nonumber\\
 &&W_4\vert_\mathrm{on-shell}=4N^2\log L+2N.\label{w4collfermi1}
 \eea
 This is a wormhole (WH), coupling the L and R copies. The term $2N$ spoils the factorization (compare to (\ref{z1fermisol})).
 \item We can have $\sigma_{LR'}=\sigma_{RL'}\equiv\sigma\neq 0$ while the other components of $\sigma_{AB}$ are zero. This is a half-wormhole. This solution reads
 \bea
 &&\sigma_{LR'}=\sigma_{RL'}=\frac{L^{N-1}}{\sqrt{2}},~\gamma_{LR'}=\gamma_{RL'}=-\imath\frac{\sqrt{2}}{L^{N-1}}\label{solz4fermi2}\nonumber\\
 &&W_4\vert_\mathrm{on-shell}=4N^2\log L+N.\label{w4collfermi2}
 \eea
 The effective action is lower than (\ref{w4collfermi1}) by $N$ so this solution is thermodynamically preferrable compared to the wormhole, and also the mismatch from factorization is smaller ($N$ compared to $2N$).
 \item The third possibility is $\sigma_{LL'}=\sigma_{RR'}\equiv\sigma$ as the only nonzero component. This solution acquires a minus sign in the logarithm of the Pfaffian (the last term in (\ref{wfermicoll})) hence it has a nonzero phase:
 \bea
 &&\sigma_{LL'}=\sigma_{RR'}=\imath\frac{L^{N-1}}{\sqrt{2}},~\gamma_{LL'}=\gamma_{RR'}=\frac{\sqrt{2}}{L^{N-1}}\label{solz4fermi3}\nonumber\\
 &&W_4\vert_\mathrm{on-shell}=4N^2\log L+\imath\pi.\label{w4collfermi3}
 \eea
 This is a half-wormhole but inequivalent to the half-wormhole from the previous point; it breaks the phase symmetry and has lower (real part of) free energy than the previous solutions. Apart from the phase factor, it is a factorizing solution, as we
 have $\Re W_4\sim 4N^2\log L=4W_1$.
 \item The last possibility is $\sigma_{LL}=\sigma_{RR}\equiv\sigma$ as the only nonzero component. This solution contributes to the last equation in (\ref{z4fermisaddle}) with the opposite sign, and thus leads to a different saddle point:
 \bea
 &&\sigma_{LL}=\sigma_{RR}=\frac{L^{N/2-2}}{2^{7/4}N^{1/4}}e^{-\imath\pi/4},~\gamma_{LL}=\gamma_{RR}=\frac{2^{7/4}N^{1/4}}{L^{N/2-2}}e^{-\imath\pi/4}\label{solz4fermi4}\\
 &&W_4\vert_\mathrm{on-shell}=4(N^2+N/2)\log L+N-\frac{1}{4}e^{-\imath\pi/4}.\label{w4collfermi4}
 \eea
 This saddle point would not exist for Majorana fermions (as in the SYK model) since for Majoranas $\bar{\psi}=\psi$ and thus $\gamma_{LL}=\langle\psi\psi\rangle=0$. It is a strongly suppressed solution as
 its on-shell action is by $\sim 2N\log L$ larger than the real part of $W_4$ of all previous solutions (\ref{w4collfermi1},\ref{w4collfermi2},\ref{w4collfermi3}).
\end{enumerate}

\begin{figure}[ht]
(A)\includegraphics[width=0.32\textwidth]{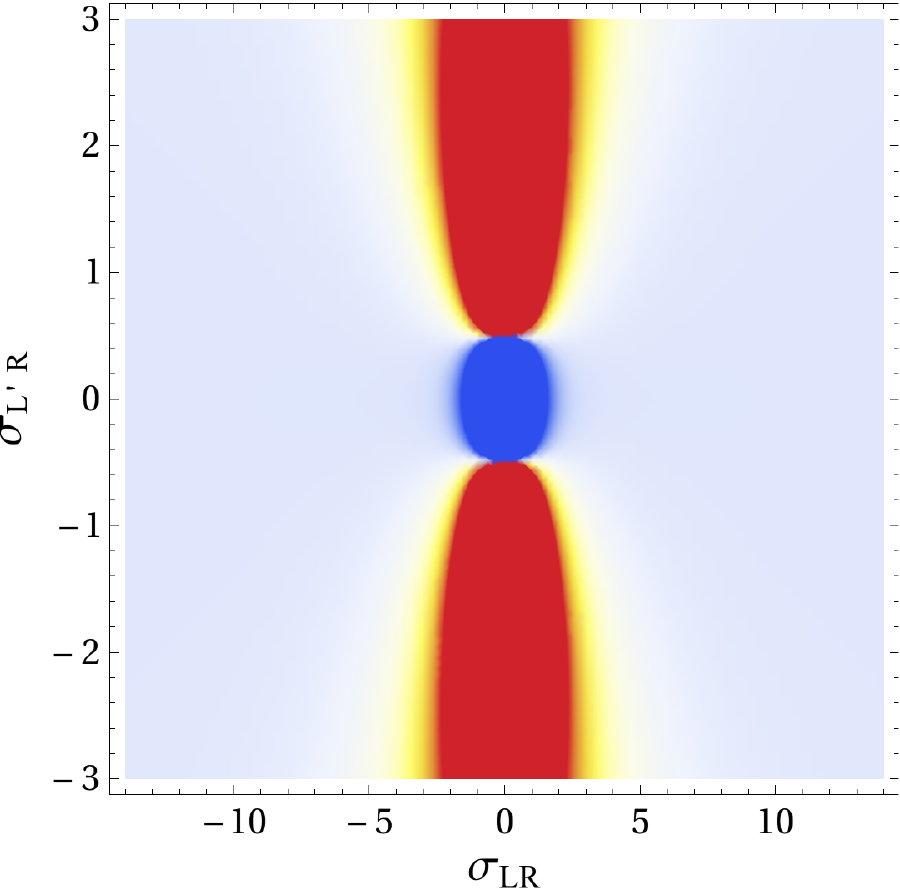}
(B)\includegraphics[width=0.48\textwidth]{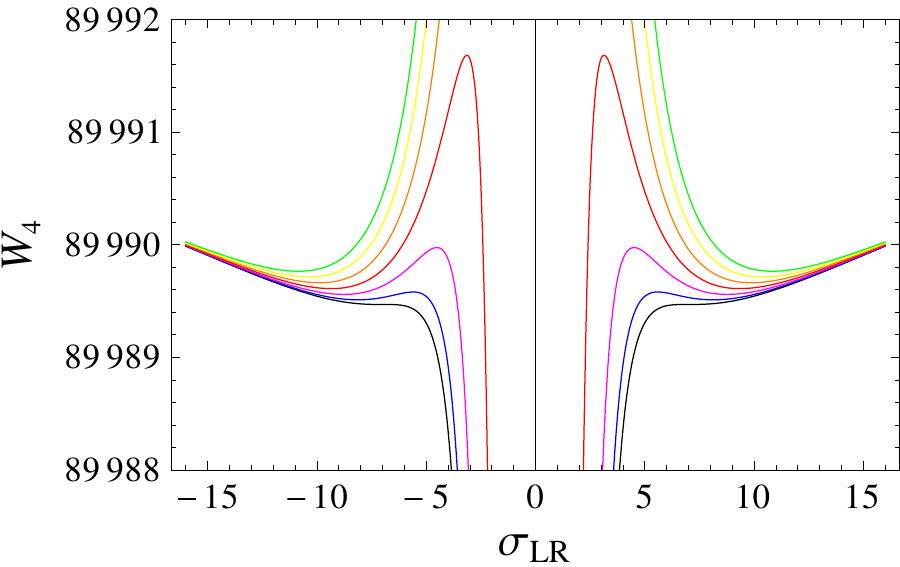}
\caption{\label{figlandscape} (A) Real part of the effective action $W_4$ as a function of the wormhole coupling $\sigma_{LR}$ and the half-wormhole coupling $\sigma_{L'R}$, for $\sigma_{LL}=\sigma_{RR}=0.3$ and $\sigma_{LL'}=\sigma_{RR'}=0.5\imath$. The 
magnitude of $W_4$ is encoded by the color map, from blue (low) to red (high). The global minimum is the trivial solution $\sigma_{LR}=\sigma_{L'R}=0$ (the blue area in the center). There is also a line of shallow local minima for nonzero WH and HWH 
couplings, which is easier to see along one-dimensional slices (B), where we plot $W_4(\sigma_{LR})$ for $\sigma_{L'R}=0.1$, $0.2$, $0.3$, $0.4$, $0.5$, $0.6$, $0.7$ (black, blue, magenta, red, orange, yellow, green). But the true vacuum, in the presence
of nonzero $LL$ and $LL'$ couplings, remains the one with no WH ($LR$) or additional HWH ($L'R$) contributions (there is already one HWH coupling $LL'$).}
\end{figure}

This situation is more akin to the SYK and similar field theory models in the literature than the purely bosonic case -- several competing solutions of the local (\ref{solz4fermi4}), wormhole (\ref{solz4fermi1}), and half-wormhole type
(\ref{solz4fermi2},\ref{solz4fermi3}). There may be more general solutions with less symmetry than the ones we found analytically above. Such solutions are hard to find, and we did not explore them in detail. In Fig.~\ref{figlandscape} we plot the
landscape of the real part of the action for varying $\sigma_{LR}$ and $\sigma_{L'R}$ (fixing the remaining couplings). The outcome is invariant to the sign change of $\sigma$ which was to be expected from the effective
action. The trivial vacuum $\sigma_{LR}=\sigma_{L'R}=\sigma_{LR'}=0$ is one solution, but we also have a nontrivial solution with all WH and HWH couplings nonzero. It seems however that the latter is never the global minimum, i.e. it is a false vacuum. So
we fail to find any thermodynamically stable solutions other than (\ref{solz4fermi1}-\ref{solz4fermi4}), but we still have no proof that they never exist.

Looking at the on-shell actions (\ref{w4collfermi1}-\ref{w4collfermi4}), we see that the dominant (least-action) solution, the $LL'$ HWH configuration, restores the factorization. It is true though that fermionic contributions are always subleading to the 
$4N^2\log L$ term coming from the bosonic determinant, so one could also argue that the factorization is always roughly satisfied; but precisely the \emph{dominant} fermionic saddle point, i.e. the true vacuum satisfies it also at the next order in $L,N$.
This strongly suggests that the physical dynamics indeed tends to "know" about the factorization. The factorization remains nontrivial, as the same dominant term $\log\det S$ from the bosonic action still gives the main contribution. And -- unlike the
findings for the SYK model in \cite{wormnonaverage} -- \emph{none of our solutions is self-averaging}. The total action for the fluctuations $a_\mu$ and $\psi_\alpha$ evaluated for $\lambda=L$ is easily found from (\ref{act2}) to be zero -- the Bose and
Fermi terms exactly cancel out each other, as it has to happen for a BPS configuration (the averaging over the quenched background $A_\mu$ spoils the supersymmetry and that is why the averaged on-shell actions we find are always nonzero; but expanding
about fixed BPS solution for $A_\mu$ without averaging keeps the BPS property). 

\section{Replicas and factorization for a pair of interacting D-strings}\label{sec4}

We have found that for a single D-string the factorization of the partition function, broken for two replicas, is always re-established with four replicas essentially at the mean field level, and in a nontrivial way. A possible physical interpretation of
this fact is simply that $n$ noninteracting D-strings form an ergodic system, i.e. averaging the dynamics over all the strings (computing their partition function $Z_n$) is equivalent to averaging a single string over the quenched degrees of freedom
(computing the partition function $\langle Z_1\rangle)$. A consequence of this is that \emph{if we have a BPS configuration, the conclusion from the previous section cannot change: as long as $F_{\mu\nu}=0$ there is no difference between $n$ copies of a
single D-brane and a stack of $n$ parallel D-branes}. Notice this remains true also in absence of background fermions (for $\Psi_\alpha=0$) since it only hinges on $F_{\mu\nu}=0$.

Now we consider a pair of strings with angle $2\theta$ between them -- this is the solution given in (\ref{stringpair}). For nonzero angle, this solution is non-BPS, the interaction between the strings is nonzero as it has a nonzero component of
$F_{\mu\nu}$. Following Eq.~(\ref{pfdefs}) and \cite{ikkt}, the superoperators are found to be:
\bea
P_0&=&\left[q\otimes I_{2\times 2},\cdot\right]=\left(\begin{matrix}I_{2\times 2}\otimes\hat{Q} & 0\\ 0 & I_{2\times 2}\otimes\hat{Q}\end{matrix}\right)\nonumber\\
P_1&=&\cos\theta\left[p\otimes I_{2\times 2},\cdot\right]=\cos\theta\left(\begin{matrix}I_{2\times 2}\otimes\hat{P} & 0\\ 0 & I_{2\times 2}\otimes\hat{P}\end{matrix}\right)\nonumber\\
P_2&=&\frac{\ell}{2}\left[I_{N'\times N'}\otimes\sigma^3,\cdot\right]=\ell\left(\begin{matrix}-\Pi_-\otimes I_{N'^2\times N'^2}  & 0\\ 0 & \Pi_+\otimes I_{N'^2\times N'^2}\end{matrix}\right)\nonumber\\
P_3&=&\sin\theta\left[p\otimes\sigma^3,\cdot\right]=2\sin\theta\left(\begin{matrix}-\Pi_-\otimes\hat{Q} & 0\\ 0 & \Pi_+\otimes\hat{Q}\end{matrix}\right)\nonumber\\
F_{03}&=&2\omega\sin\theta\left[I_{N'\times N'}\otimes\sigma^3,\cdot\right]=2\omega\sin\theta\left(\begin{matrix}-\Pi_-\otimes I_{N'^2\times N'^2} & 0\\ 0 & \Pi_+\otimes I_{N'^2\times N'^2}\end{matrix}\right),\label{stringpairfull}
\eea
and the other components are zero. For two strings, according to (\ref{stringpair}), the matrices $A_\mu$ and $a_\mu$ have the two-by-two block structure, hence the superoperators in (\ref{stringpairfull}) have the four-by-four block structure. We have 
denoted $N'\equiv N/2$ and $\Pi_\pm\equiv (I_{2\times 2}\pm\sigma^3)/2$. The superoperators $\hat{P}\equiv\left[p,\cdot\right]$ and $\hat{Q}\equiv\left[q,\cdot\right]$ are obtained by commuting with $p$ and $q$ and have the explicit form as given in
(\ref{pmularge}). Now we follow the same path as in the previous section, so we will not repeat all the technical details. Unlike the previous section, we restore the spacetime indices and write e.g. $a^\dagger_\mu a_\mu$ and not $a^\dagger a$; this is
because the different spacetime dimensions are not equivalent anymore as for a single string. The effective action reads
\bea
W_1&=&\frac{1}{2}\log\det\left[\frac{I}{L^2}+\left(1+\cos^2\theta\right)\left(a_\mu^\dagger a_\mu-I\mathrm{Tr}a_\mu^\dagger a_\mu\right)\right]+\frac{\ell^2}{2}\sin^2\theta\cdot\mathrm{Tr}a_\mu^\dagger K^2a_\mu+\nonumber\\
&+&2\omega\sin\theta\mathrm{Tr}\left(a_0^\dagger Ka_3+a_3^\dagger Ka_0\right),~~K=\left(\begin{matrix} -\Pi_- & 0\\ 0 & \Pi_+ \end{matrix}\right).\label{z1pair}
\eea
Introducing the collective fields, expanding the $\log\det$ term, and integrating out the original variables we get 
\bea
W_1&=&
\mathrm{Tr}\bigg[\frac{2\left(1+\cos^2\theta\right)}{L^{2N-2}}\left(g-a_\mu^\dagger a_\mu\right)+\frac{\ell^2}{2}\sin^2\theta\left(K^2g-a_\mu^\dagger K^2a_\mu\right)+2\omega\left(j-\sin\theta\left(a_0^\dagger Ka_3+a_3^\dagger Ka_0\right)\right)+\nonumber\\
&+&\imath s^\dagger\left(g-a_\mu^\dagger a_\mu\right)+\imath\zeta^\dagger\left(j-\sin\theta\left(a_0^\dagger Ka_3+a_3^\dagger Ka_0\right)\right)\bigg]=\nonumber\\
&=&\frac{1}{2}\log\det s^2\left(s^2-\sin^2\theta K^2\zeta^2\right)+\mathrm{Tr}\left[\frac{2\left(1+\cos^2\theta\right)}{L^{2N-2}}g+\frac{\ell^2}{2}\sin^2\theta K^2g+2\omega j+\imath s^\dagger g+\imath\zeta^\dagger j\right].~~~~~~~~\label{w1pair}
\eea
Here, $j$ is the new current, equal (on-shell) to the bilinear originating from the string-string interaction, and $\zeta$ is the corresponding auxiliary field. The two-replica action (to quadratic order) is now
\bea
W_2&=&\frac{1}{2}\log\det s_{AA}^2\left(s_{AA}^2-\sin^2\theta K^2\zeta_{AA}^2\right)+V_2\left(g_{AA}\right)+\mathrm{Tr}\left[2\omega j_{AA}+\imath s_{AA}^\dagger g_{AA}+\imath\zeta_{AA}^\dagger j_{AA}\right]\nonumber\\
V_2(g_{AA})&=&\mathrm{Tr}\left[\frac{2\left(1+\cos^2\theta\right)}{L^{2N-2}}g_{AA}+\frac{\ell^2}{2}\sin^2\theta K^2g_{AA}\right].\label{z2pair}
\eea
Finally, the four-replica solution yields
\bea 
W_4&=&\imath\mathrm{Tr}\left(s_{AA}^\dagger g_{AA}+\zeta_{AA}^\dagger j_{AA}\right)+V_2(g_{AA})+2\mathrm{Tr}\omega j_{AA}+\tilde{W}_4\nonumber\\
\tilde{W}_4&=&\frac{1}{2}\log\det\left[s_{AA}^2\left(s_{AA}^2-\sin^2\theta K^2\zeta_{AA}^2\right)+\left(A\mapsto A'\right)+S_{AAB'B'}^2\left(S_{AAB'B'}^2-\sin^2\theta K^2\zeta_{AAB'B}^2\right)\right]+\mathcal{V}\nonumber\\
\mathcal{V}&=&\frac{8}{L^{2N-4}}\mathrm{Tr}g_{AA}\mathrm{Tr}g_{B'B'}-\frac{4}{L^{2N-4}}\mathrm{Tr}G_{AAB'B'}.
\eea
The solutions to the saddle-point equations for $W_{1,2,4}$ depend crucially on whether the strings are parallel ($\theta=0$) or not ($\theta\neq 0$). We will therefore consider each case separately. Parallel strings with $\theta=0$ and thus $j=0$ reduce
to four times the result of the single-matrix calculation from the previous section; therefore we have $W_1\vert_{on-shell}=4N^2\log L$ (and analogously $W_4\vert_{on-shell}=16N^2\log L$).\footnote{The overall factor of four in these solutions compared to
the solutions (\ref{w1onshell}) and (\ref{w4onshell}) comes from the fact that we now explicitly consider the four matrices $A_\mu$, $\mu=0\ldots 3$; in Section \ref{sec3} we were working with a single matrix for simplicity.} This of course had to
happen, as this case is BPS. The old conclusion thus remains, as we have anticipated at the beginning of this section: the partition functions factorize nontrivially.

For a non-BPS configuration the picture will actually turn out to be simpler. Let us first solve the saddle-point equations for $W_1$. The outcome is
\bea
s&=&\frac{2\imath}{L^{2N-2}}\left[(1+\cos^2\theta)I+\ell^2\sin^2\theta K^2\right],~~\zeta=2\imath\omega I\\
j&=&\imath\frac{\sin^2\theta K^2\zeta}{K^2\zeta^2\sin^2\theta-s^3}\Rightarrow j\sim\frac{1}{2\omega}I\\
0&=&\frac{s^2\left(2+\imath gs-\sin^2\theta K^2\left(I+gs\right)\zeta^2\right)}{s^3-\sin^2\theta K^2s\zeta^2}\Rightarrow g\sim\frac{I}{2L^{2N-2}(1+\cos\theta^2+\ell^2\sin^2\theta K)}
\eea
This leads to the on-shell action (neglecting as usual the subleading terms):
\bea
&&W_1\vert_\mathrm{on-shell}\sim 2N^2\log L+N^2\log(4Lc_1c_2)+\ldots\nonumber\\
&&c_1=1+\cos\theta^2+\ell^2\sin^2\theta,~~c_2=\omega\sin\theta.\label{w1paircollsol}
\eea
It is easy to inspect $W_1$ and find that the leading contribution to the on-shell action value (\ref{w1paircollsol}) comes from the term with $\log\det$ in (\ref{w1pair}). Now let us look at the two- and four-replica action. These lead to very cumbersome
expressions which, however, do not present any principal difficulties so we do not write them out in full detail. The outcome reads
\bea
W_2\vert_\mathrm{on-shell}&\sim& 4N^2\log L+2N^2\log(4Lc_1c_2)+\ldots\nonumber\\
W_4\vert_\mathrm{on-shell}&\sim& 8N^2\log L+4N^2\log(4Lc_1c_2)+\ldots\label{w24paircollsol}
\eea
One can also check that nothing changes qualitatively when $\ell=0$, i.e. when the strings intersect. The limit of parallel strings is subtler: taking the limit $\theta\to 0$ directly does not make sense as it gives a divergence in $W_1$; instead one
should go back to the BPS solution (\ref{z1collsaddle}) for the collective fields, which brings the old result $W_1\vert_{on-shell}=4N^2\log L$.

Obviously, from (\ref{w24paircollsol}), the action always factorizes. But the interesting part is that the terms which contribute at (leading) order $N^2\log L$ all come from the same term $\log\det\left(s^2\ldots\right)$, obtained from the bosonic
determinant upon integrating out the microscopic variables $a_\mu$. Therefore, \emph{interacting systems have partition functions which factorize trivially}. This is in contrast to the non-interacting (and specifically BPS) configurations where
$\langle Z^4\rangle$ factorizes nontrivially, from the sum of terms $\log\det s$ and $\log\det S$ (i.e., from both two-replica and four-replica couplings). An intuitive explanation would be the following: in BPS systems, the interactions (from gravity and
from 2-form fields) precisely cancel out each other, but when multiple replicas are involved, there are combinations (like $LR$ or $LLL'L'$) where this cancelation does not happen, so the structure of the effective action is different from that of a single
replica (i.e. does not amount to four copies of a single replica). In the presence of interactions however, the dominant contribution to the free energy always comes from pairwise interactions: this is so for $W_1$, $W_2$, $W_4$ and for any $W_n$. 

\subsection{Fermionic contributions}

The fermionic contributions do not add anything fundamentally new to the picture, as they are subleading compared to the bosons. While the complete calculation is quite involved, it is easy to recognize the principle. Let us compute the fermionic
contribution to the single-replica and two-replica actions. The fermionic contribution at leading order comes from the term $\bar{\psi}_\alpha\slashed{P}\psi_\alpha$ from (\ref{act2}). Using (\ref{stringpairfull}), integrating out the quenched degrees of
freedom and then also the original fermionic fields $\psi_{\alpha}$, we get the fermionic contribution in addition to the bosonic one from (\ref{z1pair}):
\bea
W_{1f}&=&\imath\sigma^\dagger\gamma+\frac{L^2}{4}\gamma (K\otimes I_{N^2\times N^2})\gamma-\mathrm{Tr}(\gamma s^{-1}\gamma)-\log\det\left(\sigma-\frac{\ell}{2}\left(M\otimes I_{N^2\times N^2}\right)\right)\nonumber\\
M&=&I_{2\times 2}\otimes\sigma_3-\sigma_3\otimes I_{2\times 2},~~K=2I_{4\times 4}+\ell\sin\theta M,~~\gamma\sim\bar{\psi}_\alpha\psi_\alpha.\label{w1fermipaircollsol}
\eea
We remind that the matrices are now of size $2N\times 2N$, and therefore the superoperators are of size $4N^2\times 4N^2$, so they are naturally regarded as block matrices of size $4\times 4$ with each block of size $N^2\times N^2$. That is why we have the
four-by-four matrices in the expressions above. The contribution proportional to $\ell$ comes from the gap, i.e. the finite separation of the D-strings. To actually compute the saddle point solutions to (\ref{w1fermipaircollsol}) is not easy, but if we
look at the doubled system:
\bea
W_{2f}&=&W_2+\imath\mathrm{Tr}\sigma_{AB}^\dagger\gamma_{AB}+\frac{L^2}{4}\gamma_{AB'}(K\otimes I_{N^2\times N^2})\gamma_{AB'}-\mathrm{Tr}\left(\gamma_{AA}s_{AA}^{-1}\gamma_{AA}\right)-\nonumber\\
&-&\log\mathrm{Tr}\left(\tilde{\sigma}_{LR}\tilde{\sigma}_{L'R'}-\tilde{\sigma}_{LL'}\tilde{\sigma}_{RR'}+\tilde{\sigma}_{L'R}\tilde{\sigma}_{LR'}\right),~~\tilde{\sigma}_{AB'}\equiv\sigma_{AB'}-\frac{\ell}{2}M\otimes I_{N^2\times N^2},
~~~~~~\label{w2fermipaircollsol}
\eea
with $W_2$ being the bosonic contribution from (\ref{w24paircollsol}), we see that the only difference with respect to the parallel strings is the transformation $\sigma\mapsto\sigma-(\ell/2)M\otimes I_{N^2\times N^2}$. However, the determinant (and the 
trace) of the matrix $\sigma$ does not change under this transformation. This can be checked directly as $M$ is a degenerate matrix, with two zero eigenvalues. Therefore, the solutions and the on-shell values of the effective action stay the same as
before. We still have the multiple choices (\ref{w4collfermi1}-\ref{w4collfermi4}), where (\ref{w4collfermi3}), the true vacuum, actually leads to nontrivial factorization, but they all contribute only subleading terms to the bosonic action. Therefore, 
even though the outcome is mixed (trivial factorization for bosons plus nontrivial for fermions), the picture at leading order remains the same, with or without fermionic contributions -- the multi-replica solution factorizes trivially at leading order. 

To wrap up, the leading order collective field description (i.e. a mean field description) reproduces the factorization of partition functions of fluctuating D${}_p$-branes (the number of dimensions $p$ plays no role here), nontrivially for BPS-stabilized
branes and trivially for interacting branes. We will try to make a bigger picture out of this fact in the final section.

\section{Discussion and conclusions}\label{sec5}

The main outcome of our adventure can be summarized in the following way: (1) the restoration of factorization with $n\geq 4$ replicas (2) the nonfactorization of the two-replica system and the nontrivial factorization of the four-replica system for the 
noninteracting D-branes (3) the trivial factorization for any number of replicas in interacting D-brane systems (4) the absence of self-averaging. On one hand, the restoration of factorization is overall in line with the previous results from the
literature for the SYK model \cite{wormnonaverage,wormhalf,wormhalf2,wormpart,wormberry} and the general expectation that our intuition should somehow survive so that the thermodynamic potentials of multiple independent copies of any system should just add
up (i.e. the partition functions should just multiply). The points to ponder about are trivial vs. nontrivial factorization, the absence of self-averaging and above all the fact that we work with the IKKT model which is not a (fixed metric) field theory
but is itself a quantum gravity system.

The absence of self-averaging, while in itself an important difference from the SYK and similar models, is maybe less surprising than it at first appears. Self-averaging means that the physics of the averaged model can be obtained as a small correction to
a model with random fixed background (no matter which one!). That can only be true if the dynamics of the system is almost ergodic, i.e. if the system is strongly chaotic, perhaps only if it saturates the fast scrambling limit, like black holes and their
dual field theories (such as the SYK model). Therefore, it is no surprise that self-averaging is never there for more general models. This has little to do with gravity in the IIB model; it only has to do with the fact that it is not (dual to) a black
hole.\footnote{While the IKKT model should contain also the microscopic description of black holes, we don't know which background $A_\mu$ corresponds to that solution.} Thus we expect that a wide range of field theories will never be self-averaging (and
the fact that our equations can be reinterpreted as the Eguchi-Kawai discretized Yang-Mills theory apparently corroborates that). 

The trivial vs. nontrivial factorization is subtler. In the Section \ref{sec4} we have offered a somewhat handwaving explanation: the non-interacting nature of the BPS configurations is maintained by the precarious balance between different forces (in a
single system), this balance is lost in multi-replica combinations and is only restored when the contributions from all combinations are included (this is the term $\log\det S_{AAB'B'}$). We have no proof that this will happen at all orders or for all BPS
configurations. It is also not true that interacting systems necessarily have a trivial factorization. For example, the SYK model cannot in general be divided into mutually noninteracting subsystems, yet it factorizes nontrivially as found in
\cite{wormnonaverage,wormpart}, in the sense that the factorization is restored roughly as $Z^2\sim\mathrm{WH}+\mathrm{HWH}$, in other words \emph{both} a wormhole and a half-wormhole contribute significantly. This is in contrast with our backgrounds, both
BPS and non-BPS, where \emph{only} half-wormholes contribute significantly and there is no WH+HWH solution. An additional caveat is the important insight of \cite{wormpart} that different collective field descriptions are possible, leading to different 
types of HWH solutions. It seems that the trivial factorization of interacting branes, with nonzero $F_{\mu\nu}$ terms, is really a consequence of the structure of interactions in the action of the IKKT model (Eqs.~\ref{act2}-\ref{act4}). It is an
interesting task for future work to understand this better: is this a specific signature of theories with (quantum) gravity? 

To answer that question, it would be useful to look at theories where we can directly construct interesting geometries, and interpret the disorder \emph{geometrically}. This was done in \cite{tarekcoldhor}, where a system of wrapped branes on complicated
cycles within some compact manifold is represented as quiver quantum mechanics with a random superpotential. The basic strategy of our paper (modelling complicated solutions as effectively random, introducing the collective field formalism and then
computing the effective action) is quite close to what was done in \cite{tarekcoldhor}. The authors find both replica-symmetric saddles and saddles which break the replica symmetry; the latter probably give rise to non-factorizing solutions whereas the
former probably factorize.\footnote{These are just our rough guesses; neither we nor \cite{tarekcoldhor} have explicitly checked the factorization of partition functions in this setup.} In fact, as shown in \cite{tarekclass}, the same model in a different
regime gives rise to ergodicity breaking already in classical dynamics, suggestive of glassy behavior. The appealing thing is that in this case the many local minima of the free energy have an obvious interpretation: these are geometries with
multiple black holes, forming "molecules" \cite{tarekmolecules} and a direct gravity analogue of structural glasses \cite{tarekvitri}. We hope to learn more about the factorization of leading saddles in such systems in our future work.

Speaking of (non)factorization, one technical point should be emphasized. We look mainly at averaged partition functions $\langle Z^n\rangle$ and find various saddle-point solutions to the effective actions obtained as $-\log\langle Z^n\rangle$. One could 
instead look directly at the average of the logarithm: $-\langle\log Z^n\rangle$. This latter variant is called the quenched free energy in the literature on disordered systems, whereas our effective actions $W_n$ are called annealed free energies. The two
differ precisely when the system is non-factorizing, and this criterion was used to explain the non-factorization due to replica wormholes in \cite{engelhardt}. We use a slightly different criterion: direct comparison of annealed free energies for different numbers of replicas.
For our purposes either of the two approaches (comparing $\log\langle Z^n\rangle$ to $\langle\log Z^n\rangle$ versus comparing $\log\langle Z^n\rangle$ among each other for different $n$) is suitable to check the factorization, although the difference 
between the two kinds of free energy is perhaps the more usual way in statistical physics.\footnote{We thank Souvik Banerjee for pointing out this issue.}

One might worry about the fact that we always disregard the terms which scale to zero as $L,N\to\infty$, i.e. the terms where $L$ or $N$ appears with a negative power. Indeed, we have not studied the behavior of such terms and there is no way to argue
(except by performing additional calculations and checking explicitly) that these terms will obey the same factorization laws. However, even the starting ansatz for the background (with random Hermitian matrices) only makes sense for $N$ large (otherwise
we cannot speak of eigenvalue statistics), so the whole setup does not really address the regime of small $N$. Therefore, in the framework of our approach, the large-$N$ results we find are sound.

In the context of very interesting findings of \cite{gravfact}, one can also wonder if the D-branes of the IIB matrix model have a meaning analogous to the correlated bulk branes in 2D gravity models. In \cite{gravfact}, such branes act as novel 
UV degrees of freedom which, when integrated out to arrive at an IR description, introduce nonlocalities in the gravity theory; in that paper it turns out that these nonlocalities save the factorization if tuned to a specific value. Essentially the same
mechanism holds for our interacting D-branes -- their interaction is also nonlocal and guarantees factorization for any number of replicas; when the (nonlocal) interaction is absent (in the BPS case), the factorization can be violated with $n=2$ replicas,
and when restored it happens through a more complicated mechanism. We hope to reach a better understanding of this mechanism, and the relation to \cite{gravfact}, in future work.

Finally, although we were motivated to look at (non)factorization by the puzzle arising from the replica wormholes in AdS/CFT (like most of the recent papers on the subject), this question is worth asking also on its own, as it is connected to
dynamical and statistical properties of quantum gravity. Within our approach, we can in principle discuss also the matrix models of AdS geometries, with a CFT dual. One way would be to impose a background geometry along the lines of
\cite{noncomm1,noncomm2,noncomm3}, by introducing a deformation of the action (\ref{act}) to obtain matrices which satisfy the algebra of the generators of AdS isommetries. Another way is to consider a different matrix model, along the lines of
\cite{tarekvitri,tareksilico}, which corresponds to D-particles in AdS. We discuss this briefly in Appendix \ref{secappc}, however a more complete treatment is again a task for future work.

\section*{Acknowledgments}

I am grateful to K.~Hashimoto, A.~Tsuchiya, R.~Meyer and S.~Banerjee for inspiring discussions, to the anonymous referee for helpful comments and for pointing out additional relevant literature, and to the organizers of the conference "Strings and Fields
2021" for giving me the opportunity to present the first results of this work. This work has made use of the excellent Sci-Hub service. Work at the Institute of Physics is funded by the Ministry of Education, Science and Technological Development and by
the Science Fund of the Republic of Serbia, under the Key2SM project (PROMIS program, Grant No. 6066160).

\appendix

\section{Including the background fermions}\label{secappb}

Here we show that we do not lose anything by putting the background fermions to zero. Clearly this is a drastic change to the background as it destroys the supersymmetry and makes the on-shell action nonzero, but it should not strongly influence the
factorization properties. The inclusion of background fermions merely shifts the action of the background to zero, but the background is frozen and not integrated over in the quenched regime. The dynamics and partition functions are determined by the
fluctuations on top of the background, and these are not changed qualitatively when background fermions are included. Let us now show this explicitly, for a single degree of freedom (i.e., for noninteracting parallel D-strings). The outcome is that the
fermionic fluctuations become even smaller (and for $\Psi=0$, considered in the main text, they are already subleading to the bosonic ones), so our findings on factorization remain the same as those already found for bosonic fluctuations in sections 
\ref{sec3} and \ref{sec4}.

Keeping nonzero $\Psi$ and $\bar{\Psi}$, the action up to second order (we will not compute third- and fourth-order corrections) becomes:
\bea
S&=&S_1+S_2\\
S_1&=&\mathrm{Tr}\left[\bar{\Psi}_\alpha\slashed{a}\Psi_\alpha+\bar{\Psi}_\alpha\slashed{P}\psi_\alpha+\bar{\psi}_\alpha\slashed{P}\Psi_\alpha\right]\label{actpsi1}\\
S_2&=&\mathrm{Tr}\left[-\frac{1}{4}a_\mu\left(P^2\delta_{\mu\nu}+2F_{\mu\nu}\right)a_\nu+\bar{\Psi}_\alpha\slashed{a}\psi_\alpha+\bar{\psi}_\alpha\slashed{a}\Psi_\alpha+\bar{\psi}_\alpha\slashed{P}\psi_\alpha\right].\label{actpsi2}
\eea
The steps analogous to Eq.~(\ref{z1fermi}) leading to the effective action still consist of elementary Gaussian integrations, but we have a few extra terms. Comparing to the last line in (\ref{z1fermi}), we have two extra terms in the action at second
order:
\be
W_1=\imath\mathrm{Tr}\left(s^\dagger g+\sigma^\dagger\gamma\right)+V(g)+\Gamma(\gamma)-\mathrm{Tr}\left(\gamma s^{-1}\gamma\right)+\log\det\sigma+\frac{1}{4}\left(g\gamma+\gamma g\right)+\frac{1}{2}\left(\gamma\sigma^{-1}+\sigma^{-1}\gamma\right),
\ee
the last two terms being absent when $\Psi=0$. The equations of motion for $s$ are unchanged compared to (\ref{z1fermicoll}) but the others change:
\bea
&&\frac{2}{L^{2N-2}}I+\imath s+\frac{1}{2}\gamma=0,~\imath g+\frac{1}{2}\left(s^{-1}\right)^T+\gamma s^{-2}\gamma=0\nonumber\\
&&\imath\sigma+\frac{L^2}{2}\gamma-\frac{1}{4}\left(\gamma s^{-1}+s^{-1}\gamma\right)+\frac{1}{2}g+\frac{1}{2}\sigma^{-1}=0\nonumber\\
&&\imath\gamma+\left(\sigma^{-1}\right)^T-\left(\gamma\sigma^{-2}+\sigma^{-2}\gamma\right)=0.\label{z1fermicollpsi}
\eea
The relation between $s$ and $g$ is obviously unaffected, so we can express $g$ in terms of $s$ and insert in the remaining equations. The solution (ignoring the contributions which go to zero for large $N$ and $L$) reads
\be
s=\frac{4\imath}{L^{2N-2}}I\otimes E,~~g=\frac{L^{2N-2}}{8}I\otimes E,~~\sigma=-\frac{\imath L^{2N-4}}{8}I\otimes E,~~\gamma=\frac{4}{L^{2N-2}}I\otimes E.
\ee
Inserting this into the action $W_1$, we find that both the first and the second extra term in the action are negligible since they scale as $1$ and $1/L^{4N-4}$ respectively, the old fermionic terms are even more strongly subleading, and $\log\det\sigma$
doubles with respect to the on-shell value of $W_1$ found in Eq.~(\ref{z1collsaddle}). The same doubling will occur also for a multiple-replica configuration, since the relation between $\sigma$ and $\gamma$ (the fourth equation in (\ref{z4fermisaddle}))
changes only trivially when we consider multiple copies of the system: all terms get the replica indices but there are no new terms -- compare the mentioned equation in (\ref{z4fermisaddle}) and the third equation in (\ref{z1fermicoll}).

One caveat remains: as mentioned in the main text, there is no guarantee that we have found all solutions. It is thus possible that solutions with less symmetry exist which change qualitatively in the presence of background fermions. Our numerical 
explorations (Fig.~\ref{figlandscape}) suggest that this is unlikely, but clearly there is no rigorous proof.

\section{Hard vs. soft cutoff for matrix eigenvalues}\label{secappa}

In this Appendix we will rederive some of our results assuming the uniform distribution of the eigenvalues of the background matrices $A_\mu$ within some compactification interval: $-L_\mu\leq\lambda^\mu_i\leq L_\mu$. This is the usual picture in the IKKT
model \cite{ikkt,uspekhi}.\footnote{Sometimes in the literature the factor $L_\mu/\sqrt{2\pi N}$ is pulled in front of the matrices $A_\mu$, so the eigenvalues are distributed between $\pm\sqrt{2\pi L_\mu}$ and the commutators between $A_\mu$'s are just
$\imath$ with no prefactors. We have adopted the conventions of \cite{uspekhi} in this paper, where the eigenvalues are between $\pm L_\mu$ and the commutator acquires additional prefactors as in (\ref{stringpair}).} From now on we again equate all
compactification radii, so that $L_\mu\equiv L$. We will see that, apart from numerical factors of order unity, nothing changes compared to the Gaussian soft cutoff. The reason is essentially that we are looking at the quenched IKKT/large-N Yang-Mills 
model, expanding in small fluctuations $a_\mu$, $\psi_\alpha$ and in large cutoff/compactification radius $L$; it is thanks to this quenched dynamics that the system is largely insensitive on the details of the eigenvalue distribution of the semiclassical
background. If the matrices $A_\mu$ were annealed and not quenched, i.e. if we were to study the full dynamics of the background, this would not be true and we would have to be careful about the regulator. This point is explained in detail in
\cite{heckmann} where the averaged theory depends crucially on the UV closure because in the setups of \cite{heckmann} both fast and slow degrees of freedom are dynamical.

Let us start from the defining expression for the single averaged partition function (\ref{zbasic}):
\be
\langle Z\rangle=\int D[A_\mu]\int D[a_\mu]e^{-S(a_\mu;A_\mu)}\mathcal{P}(A_\mu)=\int D[a_\mu]\int_{-L_\mu}^{L_\mu} d^{2N}\lambda_\mu^ie^{-S\left(a_\mu;\lambda_\mu^i\right)},~~~~~\label{zbasicapp}
\ee
where we have now emphasized that the limits of integration for $\lambda_\mu^i$ are between $-L$ and $L$. Proceeding along the same lines as before, this yields the integral
\bea
\langle Z\rangle&=&\int D[a_\mu]\int_{-L}^L d^{2N}\lambda_{\mu i}\Pi_{i<j}\left(\lambda_{\mu i}-\lambda_{\mu j}\right)^2
\nonumber\exp\left[-\frac{1}{4}a^\dagger_{\mu ij}\left(\lambda_{\mu i}^2+\lambda_{\mu j}^2\right)a_{\mu kl}\delta_{jk}\delta_{il}\right]=\int D[a_\mu]e^{-W_1}\\
\label{z1saddleapp}W_1&=&\frac{1}{2}\sum_\mu\log\det\left(2a_\mu^\dagger a_\mu-2I\mathrm{Tr}a_\mu^\dagger a_\mu\right)-\log\mathrm{Erf}\left(L\sqrt{\frac{1}{2}\mathrm{Tr}a_\mu^\dagger a_\mu}\right).
\eea
The error function $\mathrm{Erf}$ in the result is quite difficult to work with (we understand the error function of a matrix and in general functions of matrices in the usual way). But when we expand in $a_\mu$ small just like we did in the last line of
(\ref{z1saddle}) the result simplifies:
\be
W_1=a_\mu^\dagger a_\mu-\frac{4}{3}I\mathrm{Tr}a_\mu^\dagger a_\mu+O\left(\left(a_\mu^\dagger a_\mu\right)^4\right)\mapsto\frac{1}{2}\log\det s+\imath\mathrm{Tr}s^\dagger g+\frac{2}{3L^{2N-2}}\mathrm{Tr}g.\label{z1saddlefinapp}
\ee
In the above equation, we have first performed a straightforward series expansion of the effective action $W_1$ in (\ref{z1saddleapp}), and then we have introduced the collective field $g=a_\mu^\dagger a_\mu$ just like with the Gaussian cutoff. As we see,
the additional error function terms, when expanded to quadratic order, merely change the coefficients in front of some of the terms. From here it is already obvious that the whole logic will remain the same as before; we believe there is no reason to 
repeat all the calculations again, as the factorization does not depend on the numerical coefficients (if say the term $\mathrm{Tr}g$ has a different coefficient, the same coefficient will remain also in the two-replica term $\mathrm{Tr}g_{AA}$, and the 
same holds for the potentials $V(g_{AA})$).

Now we show the same result for the fermionic fluctuations. Starting from the basic expression (\ref{z1fermi}) we can write
\bea
\langle Z\rangle&=&\int D[a_\mu]\int D[\bar{\psi}_\alpha]\int D[\psi_\alpha]\int_{-L}^L d^{2N}\lambda_{\mu i}\Pi_{i<j}\left(\lambda_{\mu i}-\lambda_{\mu j}\right)^2
\nonumber\exp\left[-\frac{1}{2}a_\mu^\dagger P^2a_\mu-\frac{1}{2}\bar{\psi}_\alpha\slashed{P}\psi_\alpha\right]=\\
&=&\int D[a_\mu]\int D[\bar{\psi}_\alpha]\int D[\psi_\alpha]e^{-W_1}\nonumber\\
e^{-W_1}&=&\sqrt{\frac{\pi}{2a_\mu^\dagger a_\mu}}e^{\frac{(\bar{\psi}_\alpha\psi_\alpha)^2}{2a_\mu^\dagger a_\mu}}
\left[\mathrm{Erf}\left(\frac{La_\mu^\dagger a_\mu+\bar{\psi}_\alpha\psi_\alpha}{\sqrt{2a_\mu^\dagger a_\mu}}\right)-\mathrm{Erf}\left(\frac{-La_\mu^\dagger a_\mu+\bar{\psi}_\alpha\psi_\alpha}{\sqrt{2a_\mu^\dagger a_\mu}}\right)\right],\label{z1fermiapp}
\eea
where it is understood that expressions of the form $(\bar{\psi}_\alpha\psi_\alpha)^2/2a_\mu^\dagger a_\mu$ really mean $(2a_\mu^\dagger a_\mu)^{-1}(\bar{\psi}_\alpha\psi_\alpha)^2$, i.e. we divide by matrices the usual way, by multiplying by the matrix
inverse. The result (\ref{z1fermiapp}) is quite involved, but a series expansion again brings it to the form quite close to the Gaussian result (\ref{z1fermi}):
\bea
W_1&=&a_\mu^\dagger a_\mu-I\mathrm{Tr}a_\mu^\dagger a_\mu+\frac{L^2}{6}\left(\bar{\psi}_\alpha\psi_\alpha\right)^2+O\left(\left(a_\mu^\dagger a_\mu\right)^4+\left(\bar{\psi}_\alpha\psi_\alpha\right)^4\right)\Rightarrow\nonumber\\
\Rightarrow W_1&=&\imath\mathrm{Tr}\left(s^\dagger g+\sigma^\dagger\gamma\right)+V(g)+\Gamma(\gamma)+\log\det\sigma,~~\Gamma(\gamma)=\frac{L^2}{6}\gamma^2,
\eea
and $V=V_2+V_4$ remains unchanged from (\ref{w2coll}). The only change with respect to the Gaussian eigenvalue statistics is the coefficient in $\Gamma(\gamma)$, being $L^2/6$ instead of $L^2/4$. Therefore, the story remains the same: while the on-shell
values of the actions $W_n$ will change, the (non)factorization properties will not, as the coefficient change in $\Gamma$ affects equally the actions $W_n$ with any number $n$ of replicas.

\section{Toward a holographic interpretation}\label{secappc}

So far, as we have mentioned in the Introduction, the connection of this work to AdS/CFT is only indirect, as the simple D-string configurations that we consider do not have an obvious CFT dual. We have also argued that it is nevertheless very relevant to
know the factorization properties of theories with gravity, precisely in order to know if the restoration of factorization through half-wormholes despite the existence of non-factorizing wormhole saddles is a specific twist of holography or rather a
generic phenomenon. But then again a realization of AdS replica wormholes, with a dual CFT, in the framework of IKKT model is clearly a worthy and exciting task. It is an open problem how to derive the holographic duality within the IKKT model, certainly
deserving a separate work; here we just outline some ideas.

Recall first that the IIB matrix model at leading order coincides with the Eguchi-Kawai model of discretized 10-dimensional Yang-Mills theory \cite{eguchikawai}. Hence, there is a direct relation to a quantum field theory: all the partition functions we 
have calculated can be reinterpreted as field-theoretical partition functions of the Eguchi-Kawai system, and these should factorize for a set of non-interacting replicas.

The second thing to note is that the IKKT model, describing the type IIB string theory, should contain also the celebrated D3 brane background which gives rise to gravity on AdS${}_5\times\mathbb{S}^5$, dual to the $\mathcal{N}=4$ superconformal Yang-Mills.
However, we do not know yet how to find such a solution explicitly in terms of matrices. A stack of parallel D3 branes is completely analogous to a pair of parallel strings that we consider, just with 4 matrices instead of 2. But the AdS geometry is not
explicit in such a solution, as the starting action and its solutions (\ref{dpbrane}) in terms of random Hermitian matrices still describe branes in flat space -- the starting action already assumes a background, because the Schild action in the first 
place also describes strings on flat background. Therefore, we could in principle repeat the averaging over some set of fluctuating AdS configurations with our formalism but it is not known what this set should be.

For this reason, there is a much studied idea \cite{noncomm1,noncomm2} to impose a geometric symmetry by deforming the model so that the matrices satisfy the commutation relations of the corresponding isommetries. For example, \cite{noncomm3} realizes the
(Lorentzian) AdS${}_2$ geometry by imposing the structure of $SO(2,1)$ algebra. In our (Euclidean) case, the group is $SO(1,2)$ and the deformed model reads:\footnote{It is not entirely clear if the fermionic part should receive any deformations, but it is
hard to think of any other meaningful term involving $\bar{\Psi},\Psi$ which satisfies the necessary symmetries. In any case, in this Appendix we will ignore the fermions completely, as we only want to provide a roadmap to a hologrpahic setup, not to
perform a precise calculation.}
\be
\label{actads}S=-\mathrm{Tr}\left(\frac{1}{4}\left[X_\mu,X_\nu\right]^2+cf_{\mu\nu\rho}\left[X_\mu,X_\nu\right]X_\rho\right),
\ee
where $f_{\mu\nu\rho}$ are the structure constants of $SO(1,2)$,\footnote{Although the commutator of classical solutions was denoted by $f_{\mu\nu}$, there can be no confusion as the structure constants have three
indices instead of two, and in this Appendix we will not make use of the commutators $f_{\mu\nu}$ anyway.} and $c$ is the coupling constant. Classical equations of motion now read (putting again the background fermions to zero):
\be
\label{eomsads}\left[X^\mu,\left[X_\mu,X_\nu\right]\right]+gf_{\nu\rho\sigma}\left[X_\rho,X_\sigma\right]=0.
\ee
But the solution to this equation has to satisfy a nontrivial commutation relation, so it is more restricted than the IKKT brane solutions, where for $N\to\infty$ \emph{any} commuting random matrices form a solution. Neverthless, for $N$ large, the three
commutation relations do not much reduce the space of solutions, and all but three eigenvalues can still be chosen at random from some interval (in other words, any reducible representation of $so(1,2)$, combined form arbitrary irreducible representations,
is a solution). The fluctuating solutions of the form $X_\mu=A_\mu+a_\mu$ now yield the following structure of the quenched action up to second order, analogous to (\ref{act2}):
\bea
S_2&=&\mathrm{Tr}\bigg[-\frac{1}{4}a_\mu\left(P^2\delta_{\mu\nu}+2F_{\mu\nu}\right)a_\nu-\imath cf_{\mu\nu\rho}\bigg(\left[a_\mu,A_\nu\right]A_\rho+\left[A_\mu,a_\nu\right]A_\rho+\left[A_\mu,A_\nu\right]a_\rho+\nonumber\\
&+&\left[a_\mu,a_\nu\right]A_\rho+\left[a_\mu,A_\nu\right]a_\rho+\left[A_\mu,a_\nu\right]a_\rho\bigg)\bigg].\label{actads2}
\eea
We will not consider the third- and fourth-order terms. For the equations of motion (\ref{eomsads}), the superoperators are expressed as $P_\mu=A_\mu\otimes I+I\otimes A_\mu$ and $F_{\mu\nu}=-f_{\mu\nu\rho}A_\rho$. Making use of that and the antisymmetry 
of the structure constants, we arrive at the simplified action for the fluctuations:
\be
S_2=-\frac{1}{4}a_\mu\left[P^2\delta_{\mu\nu}+2f_{\mu\nu\rho}\left(A_\rho-8cP_\rho-4c\Pi A_\rho\right)\right]a_\nu-2cf_{\mu\nu\rho}P_\nu A_\rho a_\mu,\label{actads2simple}
\ee
where $\Pi\equiv\sigma_3\otimes I-I\otimes\sigma_3$. Here we will be somewhat sloppy: integrating over the background $A_\mu$ requires us to know the statistical properties of eigenvalues and traces of the $so(1,2)$ representations. These are known from the
literature (see e.g. the references in \cite{noncomm2}) but the calculation goes beyond the scope of this short Appendix. For large $N$ the central limit theorems suggest that it still makes sense to assume that the eigenvalues are only weakly correlated
(which we approximate with totally uncorrelated), and limited by some cutoff function which we assume to be Gaussian, as we have shown in Appendix \ref{secappa} that the outcome is not sensitive to cutoff. In this case, paralleling the steps in the
Sections \ref{sec3} and \ref{sec4}, we find the effective action for a single copy:
\bea
W_1&=&\left(\frac{I}{2L^2}+2a_\mu^\dagger a_\mu+8c\left(a_\mu^\dagger+a_\mu\right)-2I\mathrm{Tr}a_\mu^\dagger a_\mu\right)+\frac{1}{2L^2}\left(a_\mu^\dagger f_{\mu\nu\rho}\left(1-8c-4c\Pi\right)a_\nu\right)^2\mapsto\nonumber\\
&\mapsto& W_1+\imath s^\dagger s+\imath\zeta_{\mu\nu}^\dagger K_{\mu\nu},~~g\sim a_\mu^\dagger a_\mu,~K_{\mu\nu}\sim a_\mu^\dagger a_\nu+a_\nu^\dagger a_\mu.
\eea
Therefore, because of the deformation, we have an additional set of collective fields $K_{\mu\nu}$ and the corresponding auxiliary fields $\zeta_{\mu\nu}$. Obviously, this is an interacting system, unlike the BPS parallel branes from Section \ref{sec3}. 
Integrating out $a_\mu$, we get the effective action in the form:
\bea
W_1&=&\frac{3}{2}\log\det s^2+\frac{3}{2}\log\det\zeta_{\mu\nu}^2+3\tilde{V}_2(g)-\frac{1-8c}{2L^2}\mathrm{Tr}K_{\mu\nu}+\imath s^\dagger g+\imath\zeta^\dagger_{\mu\nu}K_{\mu\nu}\nonumber\\
\tilde{V}_2(g)&=&\frac{2-32c^2}{L^{2N-2}}g.
\eea
In a system with four replicas, we get a result of the form
\bea
W_4&=&6\log\det s^2+6\log\det\zeta_{\mu\nu}^2+3\tilde{V}_2(g_{AA})+\tilde{W}_4\left(s_{AA},g_{AA},\zeta_{AAB'B;\mu\nu},K_{AAB'B';\mu\nu}\right)-\nonumber\\
&&-\frac{1-8c}{2L^2}\mathrm{Tr}K_{AA;\mu\nu}+\imath s_{AA}^\dagger g_{AA}+\imath\zeta^\dagger_{AAB'B';\mu\nu}K_{AAB'B';\mu\nu}.
\eea
We do not need detailed expressions for $\tilde{W}_4$ (although tedious, they are straightforward to find). The key point is that, just like the findings (\ref{w1paircollsol}-\ref{w24paircollsol}) for a pair of interacting D-strings, the term
$\log\det s^2$ always dominates, with the solutions $g=3\imath s^{-1}$ and $s=-(2+32c^2)/L^{2N-2}\imath I$. Finally, this yields
\be
W_n\sim -6nN^2\log L+3nN\log(2+32c^2)+\ldots,
\ee
hence we again have trivial factorization. 

A big caveat is in order: the interpretation of this and similar models is different from D-branes in the IKKT model; the bosonic matrices $X_\mu$ are not branes but quantized (non-commuting) coordinates \cite{noncomm1,noncomm2}. Besides, we have not
properly derived the eigenvalue distribution for the background. The connection to the IIB matrix model is thus rather indirect. In fact, the main limitation in applying this scheme directly to holographic setups lies in the general lack of knowledge on how
to encode AdS/CFT (i.e., the corresponding AdS backgrounds) in the IKKT action; the general idea and the formalism itself remains applicable for any background.

As a side note, another possible connection of D-brane matrix models to AdS backgrounds is the quiver matrix model of \cite{tareksilico} (see also the references therein), where D-particles in AdS are described by a supersymmetric matrix action in $0+1$
dimension, i.e. with a time derivative, hence only the static limit fits into our formalism. This setup is less directly related to the IKKT model, so we do not have a precise idea on what the meaningful background solutions would look like and what the
averaging would give.

\end{document}